\documentclass[a4paper,fleqn]{cas-sc}

\usepackage[authoryear]{natbib}
\usepackage{placeins}
\usepackage{graphicx}
\usepackage{subfig}
\usepackage{tablefootnote}
\usepackage{color}
\usepackage{multicol} 
\usepackage[pagewise]{lineno}
\usepackage{float}

\def\tsc#1{\csdef{#1}{\textsc{\lowercase{#1}}\xspace}}
\tsc{WGM}
\tsc{QE}
\tsc{EP}
\tsc{PMS}
\tsc{BEC}
\tsc{DE}

\begin{document}
\let\WriteBookmarks\relax
\def\floatpagepagefraction{1}
\def\textpagefraction{.001}

\shorttitle{\it{Cometary outbursts in the Oort cloud}}

\shortauthors{\it{Belousov \& Pavlov}}

\title [mode = title]{Cometary outbursts in the Oort cloud}

\author[1]{D. V. Belousov}[type=editor,
                        auid=000,bioid=1,
                        orcid=0000-0003-0339-5098]

\cormark[1]

\ead{dom.999.bel@gmail.com}

\affiliation[1]{organization={Ioffe Institute},
    addressline={Polytekhnicheskaya st., 26}, 
    city={St. Petersburg},
    postcode={194064}, 
    country={Russia}}

\author[1]{A. K. Pavlov}
\ead{anatoli.pavlov@mail.ioffe.ru}

\cortext[cor1]{Corresponding author}

\begin{abstract}
Comet nuclei in the outer Solar system are constantly irradiated by cosmic rays at low temperatures. Accumulated high concentrations of radicals can undergo fast recombination with significant heating of cometary surface layers. We present the model of comet activity at large heliocentric distances caused by the recombination of radicals. We found that the considered mechanism can cause activity of comets in distant regions of the Solar system, even at the Oort cloud distances. Outbursts in distant comet reservoirs can be a new source of dust and ice particles contributing to the recently discovered anomalous diffuse light in the cosmic extragalactic background optic light. The orbits of small-radii comets in the Oort cloud are highly influenced by cometary outbursts. This effect may account for the observed decrease in the number of small-radius long-period comets.       
\end{abstract}

\begin{keywords}
Comets \sep Kuiper belt\sep Cosmic rays \sep Ices
\end{keywords}

\maketitle

\section{Introduction}
\label{sec:introduction}
The outer solar system includes two sources of cometary nuclei: the Kuiper belt and the Oort cloud. The surface temperatures of comets range from 30-60 K in the Kuiper belt to about 10 K in the Oort cloud. Due to low temperatures, internal processes that comets tend to have close to the Sun (e.g., water ice sublimation) are suppressed. When considering the chemical and physical evolution of ice at low temperatures, cosmic ray irradiation plays a significant role in changing the properties of ice. In the Kuiper belt, comet nuclei are irradiated by solar cosmic rays (SCR) and galactic cosmic rays (GCR) with the modulated LIS (Local Interstellar Spectrum). The SCR flux in the Kuiper belt region is highly decreased compared to the inner part of the Solar system. In the Oort cloud, galactic cosmic rays with the LIS constantly bombard cometary surfaces. Irradiation by energetic particles generates radicals and ions in ice at low temperatures. Since the mobility of radicals decreases substantially with temperature, comet nuclei in the outer Solar system can accumulate high concentrations of radicals. Indeed, experiments on ice irradiation with protons, electrons, neutrons, and gamma-photons demonstrate efficient accumulation of radicals in ice at temperatures 10-100 K \citep{Moore1983, Carpenter1987, Shabalin2003, Zhu2021, Pavlov2022}. Radicals are highly reactive at elevated temperatures, so even a slight rise in the ambient temperature of irradiated ice can cause instant exothermic recombination of radicals \citep{Shabalin2003}. Besides, a high concentration of radicals is unstable to spontaneous recombination \citep{Shabalin2003}. When the concentration of radicals reaches 1\%, the irradiated sample can suddenly ignite due to the recombination of radicals \citep{Moore1992}.   

When approaching the Sun, a comet increases its brightness by reflecting sunlight on dust and ice particles ejected in a halo. This process can be gradual or abrupt. The latter is called a cometary outburst \citep{Gronkowski2007}. Cometary outbursts have been detected far beyond the water-ice sublimation boundary of about 3 AU \citep{Meech2009}. Amorphous-to-crystalline ice transition is supposed to be the primary source of cometary outbursts at distances from the Sun of up to 14 AU. However, recent detections of comets at much greater heliocentric distances, namely C/2017 K2 (PANSTARRS), C/2010 U3 (Boattini) and C/2014 UN271 (Bernardineli-Bernstein), demonstrate the need for other mechanisms of comet activity \citep{Bouziani2022}. In \citet{Pavlov2022}, we showed that the recombination of radicals can potentially cause comet activity at large heliocentric distances. Since recombination reactions occur at very low temperatures, comets can demonstrate activity even in the Oort cloud.  

Herein, we consider cometary outbursts in distant regions of the Solar system. To evaluate the feasibility of cometary activity in large heliocentric distances, we developed the model of cometary outbursts induced by the recombination of free radicals. The model is described in Section~\ref{sec:model description}. The results of numerical simulations are presented in Section~\ref{sec:results} and discussed in Section~\ref{sec:discussion}. The conclusions are summarized in Section~\ref{sec:conclusion}. 

\section{Model description}
\label{sec:model description}

\subsection{Basic equations}

We use models for describing processes inside a comet from \citet{Prialnik1987}, \citet{Orosei1999} and  \citet{Marboeuf2012}. All listed models were developed for comets approaching the Sun for heliocentric distances less than 14 AU. Since free radicals constantly accumulate in the upper surface layer of a comet during irradiation, we can treat radicals as a renewable source of energy inside a comet. This very feature of free radicals makes the recombination of radicals the perfect sustainable driving force for the process inside a comet at distances where solar irradiation is strongly decreased \citep{Pavlov2022}.

In our simulations, we use the following model of comet activity. The recombination of radicals increases the temperature of dust and ice in a layer. The transition from amorphous to crystalline ice starts slowly at about 120 K and proceeds quickly at 130 K \citep{Schmitt1989}. It is well-known that amorphous ice traps molecules of volatile gases (e.g., CO) at low temperatures \citep{BarNun1985}. The release of trapped gases is the accompanying process of the amorphous-to-crystalline ice transition. Owing to the high rate of recombination of radicals and large gas capacities of amorphous ice, the release of volatiles can produce high-pressure zones beneath the comet's surface. When the gas pressure exceeds the tensile strength of the comet material, the rupture of the comet layers above the high-pressure zones occurs with the ejection of dust and ice particles \citep{Gronkowski2015}.

According to current models, comets formed in cold regions with efficient condensation of volatile gases. The experimental and theoretical works show that amorphous ice dominates the comet material. From observations, CO and CO$_{2}$ gases are the most abundant components in a comet coma besides water. Therefore, our model includes four constituents of cometary nuclei: dust, water ice, CO, and CO$_{2}$. The initial ice composition is determined by the dust-to-ice mass ratio $\mu$, the mass fraction of trapped gases in amorphous ice $f_{n}$ ($n$ = CO, CO$_{2}$) and porosity $\Psi$. The mass density of a cometary layer $\rho$ is the sum of all compounds:
\begin{equation}
\label{eq:Eq1}
 \rho = \rho_{\mathrm{d}} + \rho_{\mathrm{a}} + \rho_{\mathrm{c}} + \sum_{n}(\rho_{n}^{\mathrm{v}} + \rho_{n}^{\mathrm{c}})  
\end{equation} 

where $\rho_{\mathrm{d}}$, $\rho_{\mathrm{a}}$, $\rho_{\mathrm{c}}$, $\rho_{n}$ are the mass density of dust, amorphous ice, crystalline ice, and volatiles $n$ ($\mathrm{c}$ stands for the ice state and $\mathrm{v}$ for the gas state), respectively. The effect of water vapour on the energy and local gas pressure at the temperatures considered is insignificant.

The thermodynamic evolution of the cometary nucleus obeys the global energy conservation equation:
 
\begin{equation}
\label{eq:Eq2}
\rho c \frac{\partial T}{\partial t} = \nabla(k\nabla T) - \sum_{n} c_{n} \boldsymbol{J}_{n}\nabla T - \sum_{n} H_n^{\mathrm{s}}Q_n^{\mathrm{s}} + Y_{\mathrm{ac}} + Y_{\mathrm{rec}}  
\end{equation} 

where $c$ is the specific heat capacity, and $k$ is the thermal conductivity of a cometary material. The values of $c$ and $k$ are calculated according to \citet{Marboeuf2012}. For refractories, we use the specific heat capacity of enstatite \citep{Bouziani2022}. The heat conductivity of the cometary material is strongly affected by the Hertz factor $h$ \citep{Marboeuf2012}. The typical value for $h$ lies in the 10$^{-1}$ to 10$^{-3}$ range.  

The (2) term of Eq.~\ref{eq:Eq1} describes the heat diffusion through a comet layer. The (3) term denotes the convection, where $c_n$ and $\boldsymbol{J}_{n}$ are a specific heat and a gas flux of volatiles $n$, respectively. The (4) term accounts for the energy consumption during sublimation and energy release during condensation, where $Q_n^{\mathrm{s}}$ is a gas source and $H_n^{\mathrm{s}}$ is the latent heat of $n$ gas sublimation. The (5) term describes the energy of the amorphous-to-crystalline ice transition, taking into account the energy loss due to the sublimation of trapped gases:

\begin{equation}
\label{eq:Eq3}
Y_{\mathrm{ac}} = \rho_{\mathrm{a}}\lambda \left (H^{\mathrm{cr}} - \sum_{n}f_{n}H_{n}^{\mathrm{s}}  \right ) 
\end{equation}
 
where $H^{\mathrm{cr}}$ is the latent heat of crystallization and $\lambda$ is the rate of crystallization \citep{Schmitt1989}. The (6) term accounts for the energy released during the recombination of radicals: 

\begin{equation}
\label{eq:Eq4}
Y_{\mathrm{rec}} = \sum_{x}H_{x}^{\mathrm{rec}} K_{x} {N_{x}}^{2} 
\end{equation}

where $H_{x}^{\mathrm{rec}}$ is the energy of the recombination reaction per one radical of type $x$, $N_{x}$ is the concentration of radicals and $K_{x}$ is the rate of recombination. 

The recombination of radicals obeys the bimolecular kinetic equation:

\begin{equation}
\label{eq:Eq5}
\frac{\partial N_{x}}{\partial t} = -K_{x}(T) {N_{x}}^{2} 
\end{equation}

$K_{x}$ depends on the temperature $T$ according to the Arrhenius law:

\begin{equation}
\label{eq:Eq6}
K_{x} = K_{0,x}e^{-U_{\mathrm{a},x}/k_{\mathrm{b}}T} 
\end{equation}

with the activation energy $U_{\mathrm{a},x}$ and the Boltzmann constant $k_{\mathrm{b}}$. 

The model considers radicals only in ice component of comet matter with the following reactions of radicals: 

\begin{equation}
\label{eq:Eq7}
\mathrm{H} + \mathrm{H} = \mathrm{H}_{2}
\end{equation}

\begin{equation}
\label{eq:Eq8}
\mathrm{OH} + \mathrm{OH} = \mathrm{H_{2}O_{2}}
\end{equation}

The surface boundary condition for Eq.~\ref{eq:Eq2} is:

\begin{equation}
\label{eq:Eq9}
(1-A)F_{\mathrm{b}} = \varepsilon \sigma T^{4} + k\nabla T
\end{equation}

where $A$ is the bolometric Bond albedo, $\varepsilon$ is the infrared surface emissivity and $\sigma$ is the Stefan-Boltzmann constant. The value $F_{\mathrm{b}}$ accounts for all possible sources of external heating of comet nuclei. The solar radiation dominates at distances $r_{\mathrm{h}}$ up to 10$^3$ AU. For regions beyond 10$^3$ AU (the Oort cloud), the contribution from the cosmic microwave background ($F_{\mathrm{CMB}}$), the galactic disk radiation ($F_{\mathrm{disk}}$) and the radiation of supernovae or passing luminous O, B stars ($F_{\mathrm{OB, SN}}$) should be taking into account \citep{Stern1988}, so:

\begin{equation}
\label{eq:Eq10}
F_{\mathrm{b}} = \frac{L_{\mathrm{s}}}{4\pi r{_{h}}^{2}} + F_{\mathrm{CMB}} + F_{\mathrm{disk}} + F_{\mathrm{OB, SN}}
\end{equation}

where $L_{\mathrm{s}}$ is the solar luminosity. 

Changes in the comet layer composition are obtained by solving the mass conservation equation for each component of comet matter considering amorphous-to-crystalline ice transition, gas diffusion, sublimation/condensation: 

\begin{equation}
\label{eq:Eq11}
\frac{\partial \rho_{\mathrm{c}}}{\partial t} = \lambda \rho_{\mathrm{a}}
\end{equation}  

\begin{equation}
\label{eq:Eq12}
\frac{\partial \rho_{\mathrm{a}}}{\partial t} = -\lambda \rho_{\mathrm{a}}
\end{equation}  

\begin{equation}
\label{eq:Eq13}
\frac{\partial \rho_{n}^{\mathrm{v}}}{\partial t} + \nabla\boldsymbol{J}_n = f_n \lambda \rho_{\mathrm{a}} + Q_{n}^{\mathrm{s}}
\end{equation}

\begin{equation}
\label{eq:Eq14}
\frac{\partial \rho_{n}^{\mathrm{c}}}{\partial t} = -Q_{n}^{\mathrm{s}}
\end{equation}

The gas flux of volatiles ($n$ = CO, CO$_2$) is given by:

\begin{equation}
\label{eq:Eq15}
\boldsymbol{J}_n = G_n\nabla P_{n}
\end{equation}

where $P_n$ is the partial pressure of gas $n$ and $G_n$ is the gas diffusion coefficient. The gas flow though porous comet matter can be: 1) free molecular (Knudsen regime), 2) viscous or 3) mixture of them depending on the Knudsen number. The model considers cylindrical pores with the tortuosity of $\sqrt{2}$ following \citet{Marboeuf2012}. 

The surface boundary condition for Eq.~\ref{eq:Eq13} is $P_n = 0$. The sublimation of crystalline ice is not included since it is insignificant at the considered temperatures. Diffusion of dust is also not taken into account. 

The ejection of volatiles from the amorphous ice leads to unsaturation conditions when the local pressure ($P_n$) is greater or less than the saturated gas pressure $P_n^{\mathrm{s}}$. The re-establishment of saturation conditions by means of gas condensation (when $P_n > P_n^{\mathrm{s}}$) or ice sublimation (when $P_n < P_n^{\mathrm{s}}$) occurs at much smaller timescales than the amorphous-to-crystalline ice transition. The model assumes that if the time of condensation/sublimation is less than the chosen time-step of the model, the re-establishment of saturation conditions occurs instantly. The deviation of the local pressure from the saturated pressure is added/subtracted to/from the density of the condensed gas to guarantee mass conservation \citep{Davidsson2021}. 

The release of trapped gases from the amorphous ice results in gas pressure growth. The pressure continues increasing until the gas flux from amorphous ice reaches the balance with the diffusion flux casting through the comet layer. The values for the pressure in subsurface areas can be calculated using Eq.~\ref{eq:Eq2}-\ref{eq:Eq15}. When subsurface pressure exceeds the tensile strength of the comet matter, fractures can develop. This effect may lead to the intense ejection of comet matter over cavities with high gas pressure \citep{Gronkowski2015, Prialnik2017}. Here, we assume that the outburst goes off when subsurface pressure exceeds the tensile strength.

In the numerical calculations, we use a fully implicit scheme with a finite volume method for Eq.~\ref{eq:Eq2} and \ref{eq:Eq13} in 1D spherical geometry. The time step is constantly adjusted during calculations in order to restrict changes in pressure and temperature.

\subsection{Initial conditions and model parameters}

Here, we consider comet nuclei located in the Oort cloud. Although the composition of comets in the Oort cloud is unknown, observed dynamically "new" comets may represent comets in the Oort cloud. Since the model simulates the evolution of upper layers of comets, where effective irradiation by cosmic rays occurs, all the parameters listed hereafter describe the surface layers. Initially, the volatile gases CO and CO$_2$ are present only as trapped in the amorphous ice. We use this constraint to distinguish this model from the model of cometary activity at large heliocentric distances based on the sublimation of volatile ice. In addition, the sublimation of condensed and trapped gases has a similar effect on the total energy of cometary activity (see Eq.~\ref{eq:Eq2}). The molar abundances of the gases CO and CO$_2$ relative to water are both 10 $\%$ (i.e., the mass ratios of CO and CO$_2$ to water ice are 0.16 and 0.25, respectively). We chose these values for $f_{\mathrm{CO}}$ and $f_{\mathrm{CO_2}}$ for two reasons: 1) trapped volatile molecules in high concentrations make the amorphous-to-crystalline ice transition energy neutral. Thus, recombination reactions are the main mechanism of comet activity in this model; 2) the molar abundance of interstellar CO$_2$ ices is in the range of $10 - 23 \%$ with approximately the same value for CO \citep{Gerakines1999}. In the model dust to ice mass ratio varies from the ice-rich $\mu = 0$ to the ice-poor $\mu = 4$ cases. The pore radius is set to 1 $\mu$m following \citet{Gronkowski2015}. The intense GCR irradiation in the upper comet layer can produce a dense crust enriched with organic molecules and dust \citep{Strazzulla1991}. Here, we assume the dense irradiation-produced crust is made of refractories (organics and dust) with low porosity $\Psi = 0.1$ and a density close to 3000 kg m$^{-3}$. Under the crust, comet matter is supposed to be very porous ($\Psi$ = 0.65). All parameters used in simulations are listed in Table~\ref{tab:parameters}. The observations of the 67P/Churyumov–Gerasimenko show that comets tend to be low-strength bodies with tensile strength as low as a few Pa \citep{Basilevsky2016}. However, the large-scale structure of long-period comets can significantly differ from that of short-period comets. In addition, the tensile strength could depend on the depth of a comet layer. In simulations, we use the maximum estimated value of the large-scale tensile strength  of comet matter 10$^4$ Pa \citep{Reach2010} following \citet{Gronkowski2015}.  

Free radicals can be activated: 1) when the ambient temperature increases (induced recombination) or 2) when a critical concentration of radicals is reached (spontaneous recombination). The model considers only induced recombination. The activation energy of H radicals (Eq.~\ref{eq:Eq6}) varies from 0.01 eV to 0.11 eV \citep{Flournov1962, Kirichek2017}. At temperatures above 50 K, H radicals are present only in trace amounts compared to hydroxyl radicals. OH radicals are stable at low temperatures and react violently at about 100 K with an activation energy of 0.25 eV \citep{Johnson1997, Pavlov2022}. The recombination of hydroxyl radicals starts at the temperature of 100 K, so the rate of recombination at temperatures below 100 K is negligible. Thus, we do not consider the mutual recombination of OH and H radicals in our model; the parameters of this reaction are poorly known from the experiments. According to the model, the comet layer must be heated by at least one kelvin to induce the recombination of H radicals. Parameters for Eq.~\ref{eq:Eq6} are from \citep{Pavlov2022} and \citep{Flournov1962}. Long-term irradiation at low temperatures leads to the accumulation of high radiation doses in surface layers of comets. GCRs (mainly protons) effectively deposit energy in the upper 10 meters of the comet's surface. \citet {Gronoff2020} modelled the irradiation of a comet ($\rho$ = 538 kg m$^{-3}$, $\mu$ = 4) by GCR and obtained the absorbed dose profile. A particular layer of a comet is able to accumulate radicals with a concentration not exceeding the critical value. From experiments, spontaneous recombination in the proton-irradiated water ice occurs at the concentration of H radicals equal to about 1$\%$ (to the number of water ice molecules)\citep{Moore1992}. Using the yield of H radicals in the proton-irradiated amorphous ice 7$\times$10$^{-3}$ eV$^{-1}$ from \citet{Moore1992} and the deposition dose rate in the Oort cloud from \citep{Gronoff2020}, the time of the critical concentration accumulation varies from 10$^2$ Myr at 1 meter to 10$^3$ Myr at 10 meters beneath the comet surface. The first centimetres of a cometary nucleus reach the critical concentration of radicals within the 5 Myr period. This effect may favour the formation of a dense crust.  

\begin{figure}[h]
	\centering
		\includegraphics[scale=.17]{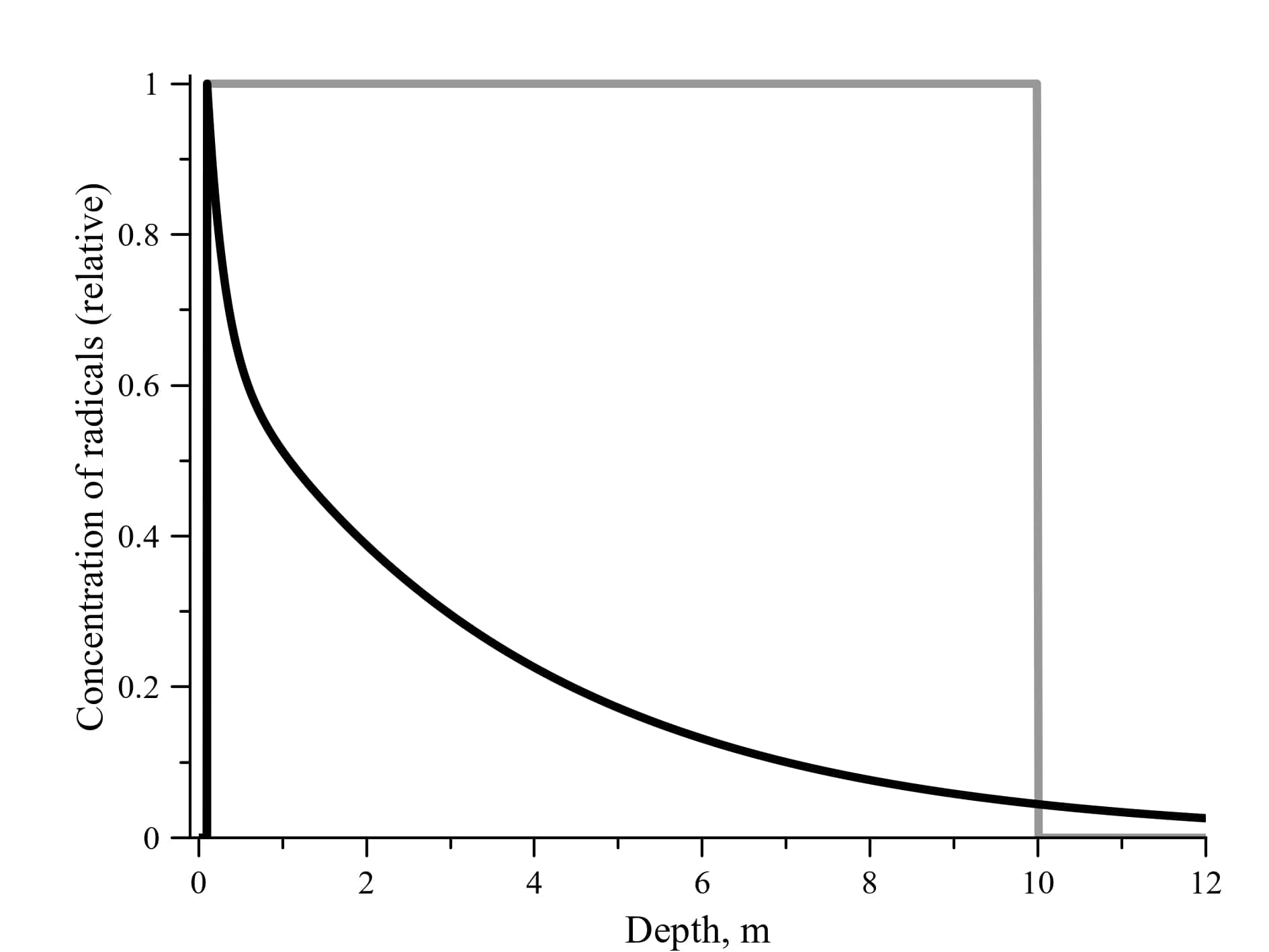}
	\caption{\rm{Sample distributions of radicals in the comet located in the Oort cloud. The black solid curve represents the case of the maximum concentration beneath the surface dust layer (1). The grey solid curve shows the distribution of radicals when radicals reach the saturated value (2). The dose deposition rates are from \citet{Gronoff2020}.}}
	\label{FIG:RadicalsDistribution}
\end{figure}

In the model, we use two profiles of radicals distribution. The first profile corresponds to the moment when the high concentration of radicals is achieved directly under the dust layer. The second profile corresponds to the accumulation of equilibrium saturated concentration of radicals in surface layers (Fig.~\ref{FIG:RadicalsDistribution}). According to \citet{Shabalin1998}, the saturated concentration of radicals can be independent of the irradiation dose rate at low temperatures. As a result, the upper cometary layers can have a similar saturated concentration of radicals. Here, we assume that only the upper 10 meters of the comet nucleus accumulate a high concentration of radicals during the evolution of the Oort cloud based on the dose deposition rates \citep{Gronoff2020}. We propose that the initial temperature gradient in the upper 10-meter layer is insignificant. The concentrations of radicals are close to or less than the critical concentration ($\approx 1\%$). The upper 10 cm layer is a dust layer without radicals. 

Since the model considers the induced recombination of radicals, comet surface layers must experience external heating to provoke recombination reactions. In extremely distant regions (i.e., the Oort cloud), the impact between comets and the variation in solar radiation during the orbital motion is insignificant. However, close supernovae or a passing O/B star can heat cometary nuclei in the Oort cloud \citep{Stern1988}. \citet{Stern1988}  showed that the passage of very bright stars (classes O and B) in the vicinity of the Oort cloud during the evolution of the Solar system $\approx$ 4.5 Byr resulted in at least one heating event, which caused the temperature of each cometary nucleus to rise to 16 K. Also, close supernovae cause periodic heating to 30 K for all comets in the Oort cloud. The layer with radicals must be heated by at least 1 K for $U_{\mathrm{a, H}}$ = 0.01 eV or reach the temperature T = 30 K for $U_{\mathrm{a, H}}$ = 0.11 eV in order to induce the recombination of H radicals. 

\begin{table}[H]
  \caption{\rm{Parameters of the model.}}
  \begin{tabular*}{\tblwidth}{@{} LLLL@{} }
   \toprule
    \rm{Parameter} & \rm{Value}\\
   \midrule
    \rm {Comet radius R, m} & $10^{3}$\\
    $\Psi$ & $0.65$\\
    $r_{p}$, m & $10^{-6}$\\
    $h$ & $10^{-3} - 0.1$\\
    $\epsilon$ & $0.96$\\
    $A$ & $0.04$\\
    $f_{CO}, f_{CO_{2}}$ & $0.16, 0.25$\\
    \rm{Initial temperature} $T_{0}$, K & $10$\\
    \rm{Initial concentrations of radicals} $N_{0,x}$ & $0 - 1\%$\\
    $U_{\mathrm{a, H}}$, \rm{eV} & $0.1-0.11$ \\
    $U_{\mathrm{a, OH}}$, \rm{eV} & $0.25$ \\
   \bottomrule
  \end{tabular*}
  \label{tab:parameters}
\end{table}

\section{Results}
\label{sec:results}
\subsection{Pressure building}

Numerical calculations have been carried out using the wide range of possible values for the Hertz factor $h$, the dust-to-ice mass ratio $\mu$, the activation energy of H radicals $U_{\mathrm{a, H}}$ (Table~\ref{tab:parameters}) and implementing two different spatial distributions of radicals (Fig.~\ref{FIG:RadicalsDistribution}).

\begin{figure}[h]
	\centering
		\includegraphics[scale=.3]{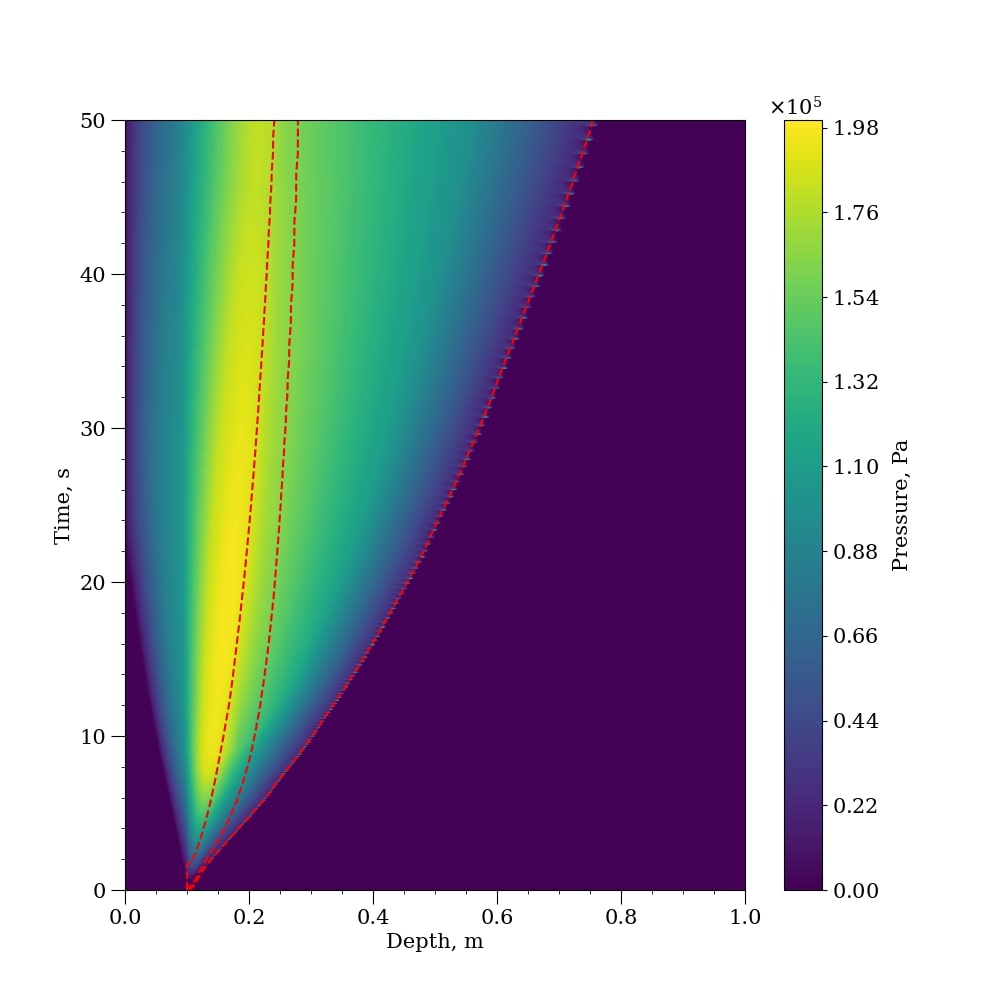}
	\caption{\rm{Pressure building inside a comet caused by the recombination of radicals. Red dashed curves depict (from left to right) positions of the amorphous-to-crystalline transition, the OH and H radicals’ recombination fronts. Calculations were conducted using the first distribution of radicals (see Fig.~\ref{FIG:RadicalsDistribution}) with the maximum concentration of 1$\%$, parameters from Table~\ref{tab:parameters} and $h$ = 0.1, $\mu$ = 1, $U_{\mathrm{a, H}}$ = 0.03 eV. Here and further in all figures, the time of the outburst is counted with the onset of the recombination of H radicals.}}
	\label{FIG:0.03eV_1dist}
\end{figure}

We first examine the comet nucleus with the first distribution of radicals. The pressure distribution under the comet surface during the comet activity caused by the recombination of radicals is shown in Fig.~\ref{FIG:0.03eV_1dist}. Here, the maximum concentration of radicals is 1$\%$ (relative to the number of water ice molecules; the first distribution Fig.~\ref{FIG:RadicalsDistribution}). Our simulations show that the efficient pressure growth occurs near the source of gas – the amorphous-to-crystalline ice transition front. In addition, effective sublimation and condensation of CO gas take place at the H recombination front with the concomitant local gas pressure building (Fig.~\ref{FIG:0.03eV_1dist}). In the model, we assume that the ice transition front position determines the depth where the initial concentration of amorphous ice decreased by 10$\%$. Similarly, positions of H and OH recombination fronts are determined by a local drop in the concentration of radicals by 10$\%$. According to Fig.~\ref{FIG:0.03eV_1dist}, the gas pressure accumulated below the comet crust exceeds the tensile strength of comet material ($\approx$ 10$^4$ Pa), so the comet crust and all material above cavities with the high gas pressure can be ejected. 

\FloatBarrier 

\begin{figure}[h]
	\centering
		\includegraphics[scale=.17]{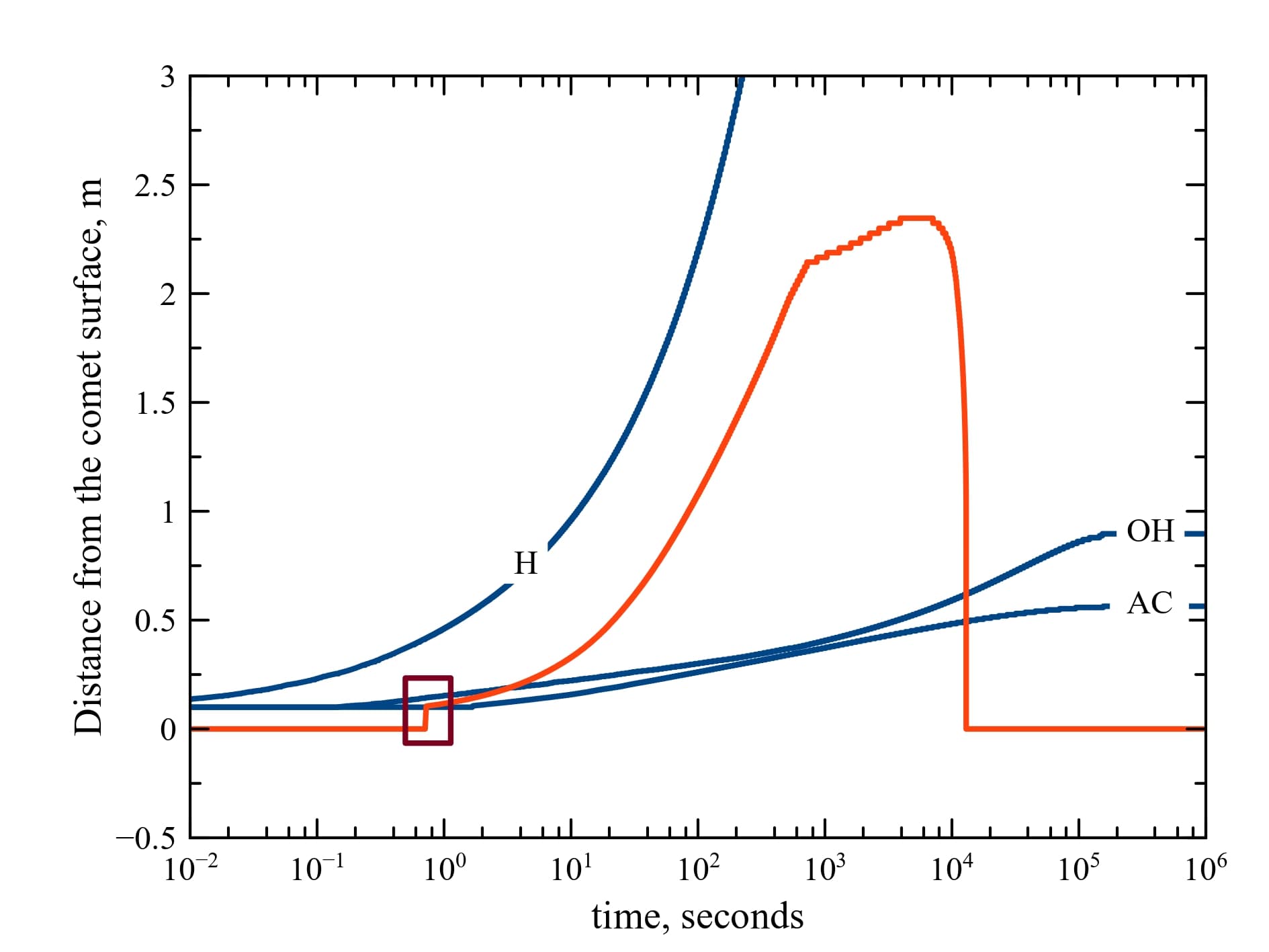}
	\caption{\rm{Physical evolution of a comet during the outburst. The positions of amorphous-to-crystalline transition (AC) and OH, H radicals' recombination fronts are presented by blue curves. The positions of the deepest layers with the gas pressure > 10$^4$ Pa are shown by the red curve. The purple square depicts the initial rise in gas pressure coinciding with the ice transition front. Initial conditions and parameters are as in Fig.~\ref{FIG:0.03eV_1dist}, except $U_{\mathrm{a, H}}$ = 0.01 eV.}}
	\label{FIG:Fronts_1Distribution}
\end{figure}

The propagation of recombination and amorphous-to-crystalline transition fronts, as well as the location of zones with gas pressure exceeding the tensile strength, are shown in Fig.~\ref{FIG:Fronts_1Distribution}. The initial pressure rise takes place at the ice transition front within a few seconds after the onset of H recombination. Further propagation of volatiles is governed by effective gas diffusion owing to deeper layers heating by H and OH radicals. According to Fig.~\ref{FIG:Fronts_1Distribution}, only 0.5 meters of a comet nucleus undergoes an ice transition, while high-pressure zones can extend up to about 2.5 meters. 

\FloatBarrier 

\begin{figure}[h]
	\centering
		\includegraphics[scale=.17]{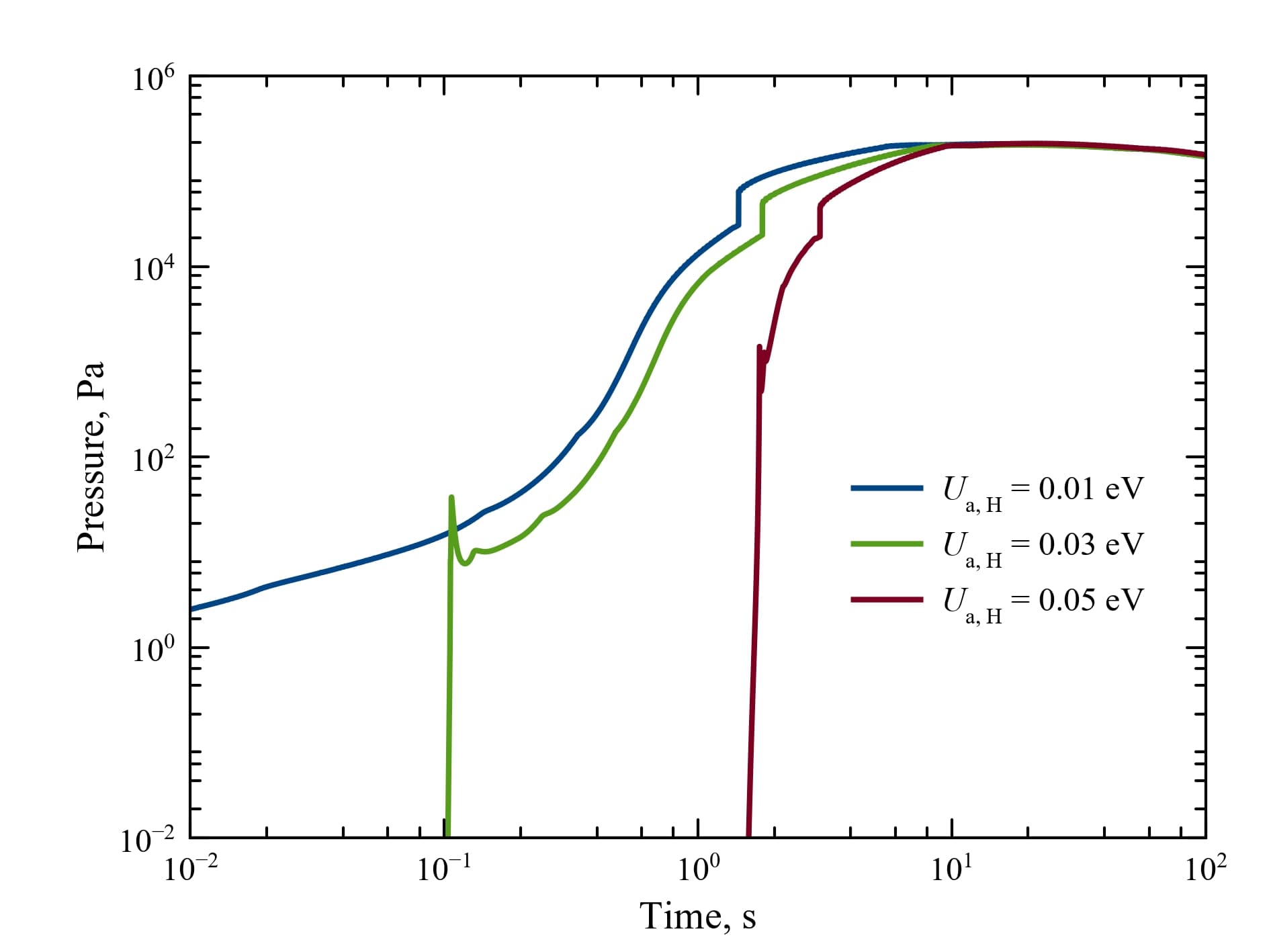}
	\caption{\rm{Pressure building at the amorphous-to-crystalline ice transition front as a function of time for different activation energies of H radicals. The initial conditions and parameters of simulations are as in Fig.~\ref{FIG:0.03eV_1dist}, but various $U_{\mathrm{a, H}}$.}}
	\label{FIG:Pressure_Uact}
\end{figure}

\begin{figure}[h]
	\centering
		\includegraphics[scale=.17]{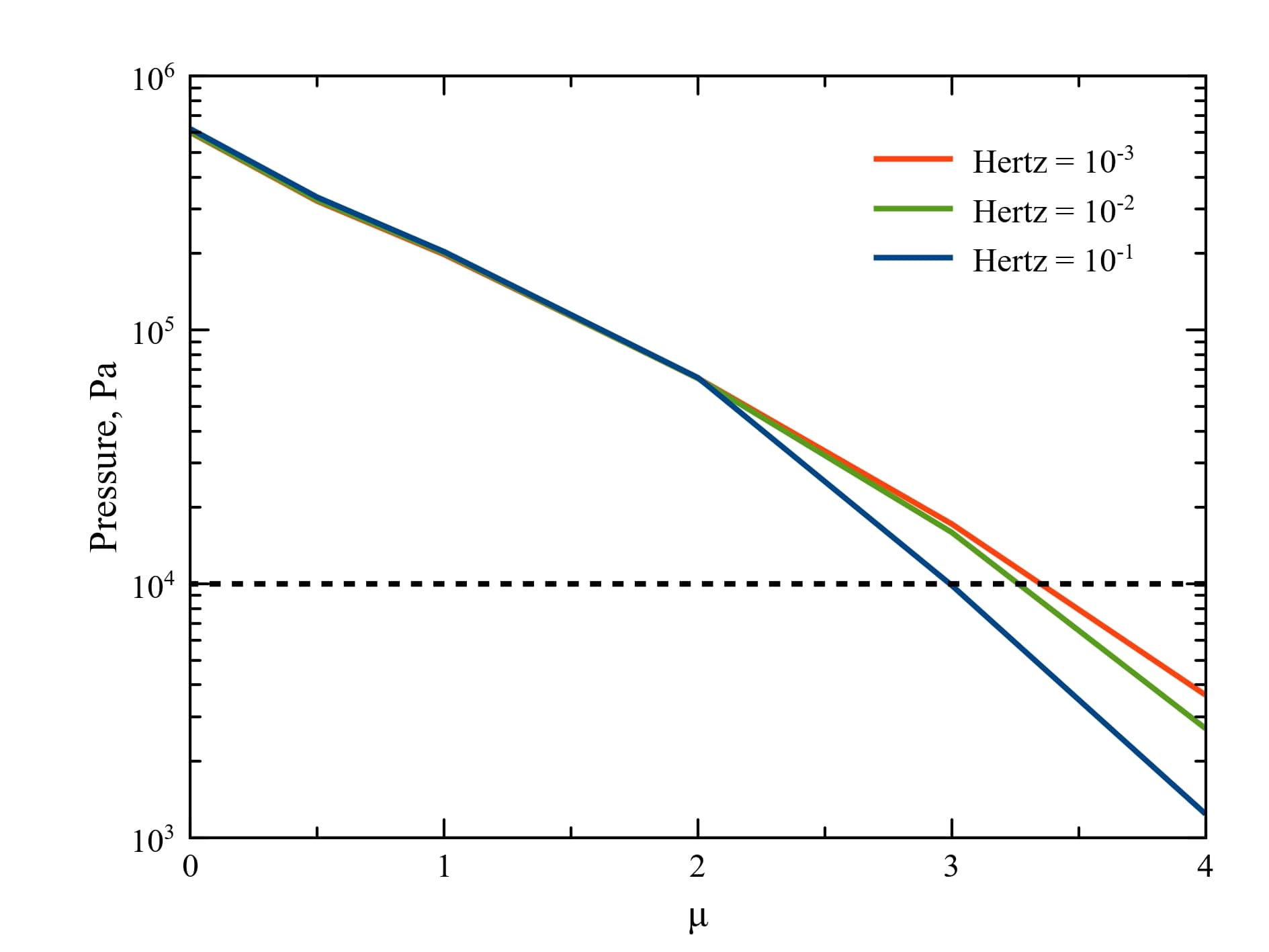}
	\caption{\rm{Maximum gas pressure accumulated under the surface of a comet, depending on the dust-to-ice ratio $\mu$ for three values of the Hertz factor. The black dashed line represents the tensile strength of the comet material \citep{Reach2010}. The initial conditions and parameters of simulations are as in Fig.~\ref{FIG:0.03eV_1dist}, but for $U_{\mathrm{a, H}}$ = 0.01 eV and various $h$ and $\mu$.}}
	\label{FIG:Pressure_HertzFactor}
\end{figure}

The total pressure of CO and CO$_2$ gases achieves its maximum value within 10 – 100 seconds (Fig.~\ref{FIG:Pressure_Uact}). We found that the maximum gas pressure depends slightly on the activation energy of H radicals (see Fig.~\ref{FIG:Pressure_Uact}). The pressure values accumulated in subsurface cavities depending on the dust-to-ice mass ratio $\mu$ and various values of the Hertz factor $h$ are presented in Fig.~\ref{FIG:Pressure_HertzFactor}. Based on a comparison with the tensile strength of comet matter, we conclude that the surface layers of comets with $\mu = 0 - 3$ and a high concentration of radicals can experience outbursts. This effect is slightly influenced by the Hertz factor (Fig.~\ref{FIG:Pressure_HertzFactor}). 

\FloatBarrier 

\FloatBarrier 

\begin{figure}[h]
	\centering
		\includegraphics[scale=.17]{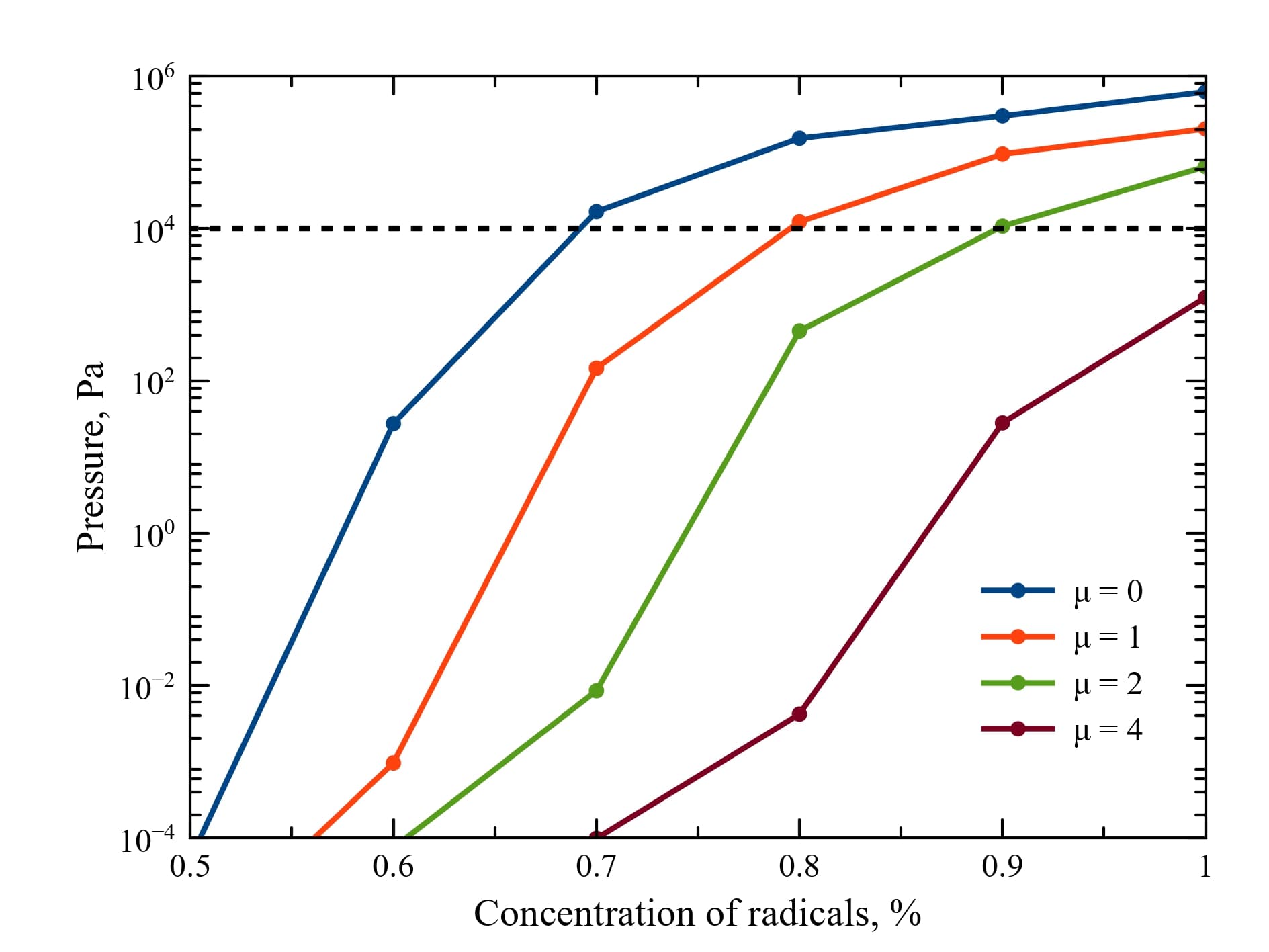}
	\caption{\rm{Maximum gas pressure accumulated under the surface of a comet as a function of maximum concentration of radicals in ice (the first distribution, see Fig.~\ref{FIG:RadicalsDistribution}), for four values of the dust-to-ice mass ratio $\mu$. The black dashed line represents the tensile strength of the comet material \citep{Reach2010}. Calculations were carried out using parameters from Table~\ref{tab:parameters} and $h$ = 0.1, $U_{\mathrm{a, H}}$ = 0.01 eV.}}
	\label{FIG:Pressure_1Dist}
\end{figure}

Concentrations of radicals sufficient to cause the significant pressure building in the subsurface layers are presented in Fig.~\ref{FIG:Pressure_1Dist}. To achieve gas pressure exceeding the tensile strength, the upper comet layer (just below the dust layer) must accumulate a concentration of radicals in the ice component above 0.7 $\%$.

\FloatBarrier 

\begin{figure}[h]
	\centering
		\includegraphics[scale=.3]{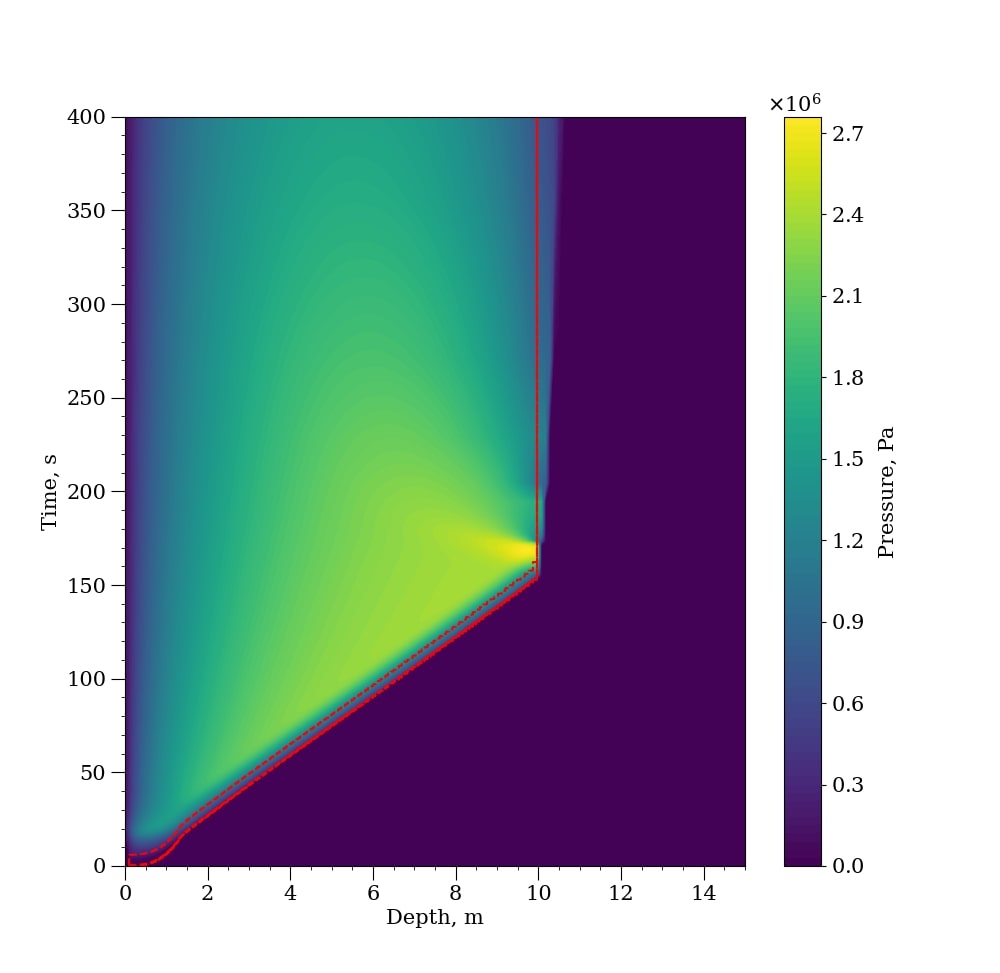}
	\caption{\rm{As Fig.~\ref{FIG:0.03eV_1dist}, but for the second distribution of radicals (see Fig.~\ref{FIG:RadicalsDistribution}) with the saturation concentration of 1$\%$, $h$ = 0.1, $\mu$ = 1 and $U_{\mathrm{a, H}}$ = 0.01 eV. The positions of H and OH recombination fronts coincide on this plot.}}
	\label{FIG:0.01eV_2dist}
\end{figure}

We now consider the second distribution of radicals. We find that the effective pressure building follows the recombination fronts (Fig.~\ref{FIG:0.01eV_2dist}). Due to the low activation energy of H radicals, the recombination front of H radicals effectively casts through a comet, simultaneously heating cometary layers to approximately the same temperature (since we use the uniform saturated concentration of radicals). If the concentration of radicals is enough to induce ice transition, all layers undergoing recombination of radicals can accumulate high gas pressure. Calculations show that the total pressure rise depends slightly on the activation energy of H radicals. According to Fig.~\ref{FIG:0.01eV_2dist}, diffusion of volatiles results in the accumulation of high gas pressure in layers lying deeper than the recombination fronts.

\FloatBarrier

\begin{figure}[h]
	\centering
		\includegraphics[scale=.17]{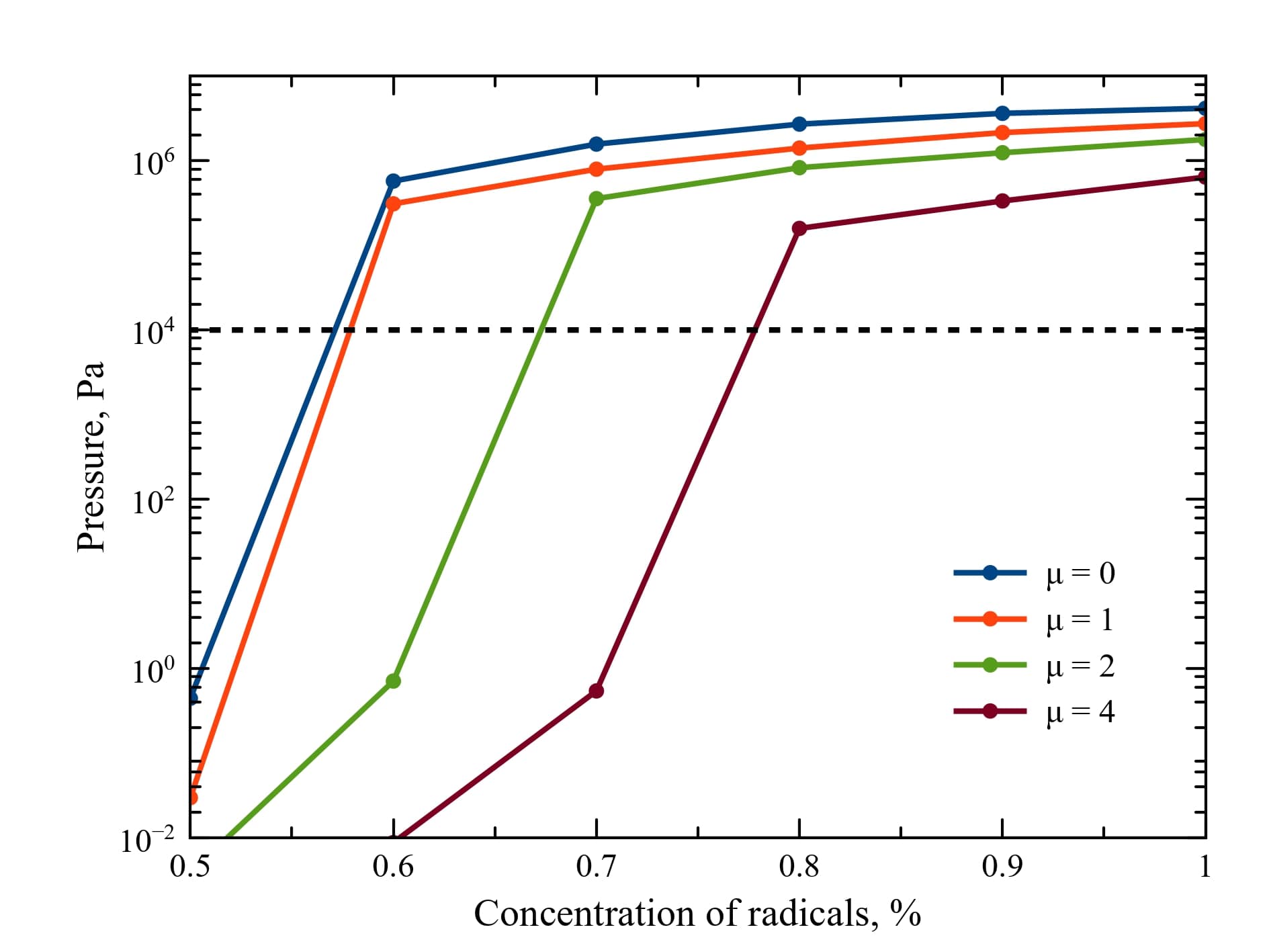}
	\caption{\rm{As Fig.~\ref{FIG:Pressure_1Dist}, but for the second distribution of radicals (see Fig.~\ref{FIG:RadicalsDistribution}).}}
	\label{FIG:Pressure_2Dist}
\end{figure}

From Fig.~\ref{FIG:Pressure_2Dist}, comets with the saturation concentration of radicals above 0.6 $\%$ (to the number of water ice molecules) are surface layers that can experience burst activity. Compared to the first distribution of radicals, recombination of high saturation concentration of radicals creates higher pressure in subsurface cavities and causes instabilities in cometary layers with $\mu = 0 - 4$. 

\FloatBarrier

\subsection{Ejected mass}
The main goal of the present study is to assess the efficiency of the recombination reaction as a possible mechanism of cometary outbursts at large heliocentric distances. So, the valuable outcome of the model is the mass loss of a comet nucleus at the Oort cloud distances. 

Based on the model of \citet{Ipatov2012} and \citet{Gronkowski2015}, when local gas pressure exceeds the tensile strength of a cavity, instabilities can develop, resulting in the formation of cracks or even the total removal of subsurface layers into a comet coma. To eject layers, the gas pressure must exceed the sum of the weight of layers above the cavity and the tensile strength. According to our results, cavities with a gas pressure high enough to reject surface layers can be formed up to $\approx$ 10 meters below the surface caused by the induced recombination of radicals. The active surface area of a comet experiencing an outburst can be obtained from observations. \citet{Kelley2022} found that the ejected mass during the outburst of long-period comet C/2014 UN271 (Bernardinelli-Bernstein) at a distance of 20 AU originated from the cylindrical volume with a diameter of about 130 m. This value is comparable to the diameters of observed pits on the surface of short-period comets, such as 67P/Churyumov-Gerasimenko, 17P/Holmes \citep{Vincent2015, Reach2010}. The thickness of cometary layer experiencing activity is determined by the maximum depths of subsurface layers with the high gas
pressure: $\approx$ 2 and 10 m for the first and second distributions of radicals, respectively. With a dust layer density of 3000 kg m$^{-3}$ and an average density of comet matter of 500 kg m$^{-3}$ ($\mu$ = 1), the ejected mass is equal $(0.7 - 2.7)\times 10^9$ kg. This value lies within the mass range of observed cometary outbursts \citep{Gronkowski2015}.

\section{Discussion}
\label{sec:discussion}

\subsection{A possible source of distant comet activity}
We have shown that comet nuclei in distant regions of the Solar system can become active due to the recombination of free radicals accumulated in surface layers under cosmic ray irradiation. For a wide range of possible parameters for comet matter composition, the gas pressure in subsurface cavities overcomes the large-scale tensile strength. The main contribution to the gas pressure is CO gas ejected from the amorphous ice. High gas pressure creates instabilities and even causes the total rejection of surface layers. In addition, gases ejected from the amorphous ice can flow both to the surface of a comet and diffuse into deeper layers (Fig.~\ref{FIG:0.03eV_1dist}, \ref{FIG:0.01eV_2dist}). Local heating due to the recombination of radicals manages efficient gas diffusion and prevents the instant condensation of gases. This effect leads to the enrichment of underlying layers with volatiles. Recombination of radicals could be one of possible sources of the observed activity of long-period comets (C/2017 K2 (PANSTARRS), C/2010 U3 (Boattini) and C/2014 UN271 (Bernardineli-Bernstein)) at large heliocentric distances. Our model can be used to simulate the surface activity of long-period comets, Kuiper belt objects and Trans-Neptunian objects. An important indicator that recombination reactions do occur on the surface of small bodies in the outer Solar system should be the presence of substances such as H$_2$O$_2$ and O$_2$ \citep{Johnson1997}. The James Webb Space Telescope can be utilized to detect radiation products on the surface of small bodies in the outer Solar system.

\subsection{Model constrains}

In our calculations, we assume that radicals can be accumulated only in the ice fraction of the comet matter. However, radicals can be also produced in the refractories (dust and organics) during the irradiation. The accumulation of radicals in organic molecules was discovered under proton, alpha and neutron irradiation at low temperatures \citep{Carpenter1987, Kaiser1997, Shabalin2003}. The critical concentration of H atoms in methane molecules is $6\pm 3\%$ \citep{Kaiser1997}. On the other hand, there are no experimental results on the accumulation of radicals in dust. Since the mass fraction of dust in the comet matter can be comparable to or even overcome the fraction of water ice, radicals in dust can contribute substantially to the processes on cometary surfaces. It is well known that the decay of radionuclides continuously irradiates the inner layers of comets, generating radicals in much deeper layers than considered in our model. We expect changes in chemical and physical composition due to the recombination of accumulated radicals in the entire volume of a comet. It is worth mentioning that the accumulation and recombination of radicals, as well as values of critical and saturation concentrations depend on the ice matrix \citep{Johnson1997}. There are no experimental results on the accumulation of radicals in cometary analogues. So, in our calculations, we used experimental results for pure ice. Nevertheless, experiments are desirable to simulate the irradiation of cometary matter.  

From observations, comets are rather heterogeneous bodies with chemical compositions and physical conditions changing from place to place. As a result, comets can experience activity in different areas where ice with radicals meets conditions to start chain recombination reactions. We saw in simulations that even a slight rise in ambient temperature can start propagation of the recombination front, so it is crucial to properly model the heating of the surface, taking into account all surface irregularities.  

In our model, we used absorbed doses of GCRs from \citep{Gronoff2020}. It should be pointed out that \citet{Gronoff2020} used in calculations LIS based on the Badhwar–O'Neill model. The flux of low-energy particles used in \citet{Gronoff2020} are greatly overestimated compared to \citet{Boschini2020}, who used the GalProp–HelMod model and several instruments, including Voyager 1, to constrain LIS. We estimated that the new LIS would decrease the dose rate only in the upper 10 cm of a comet. In addition, GCR spectra could change during the Solar system evolution, owing to supernovae and the early Sun's activity \citep{Gronoff2020}. 

\subsection{Astronomical implication}

The studied mechanism of comet activity driven by the recombination of radicals substantially increases the range of heliocentric distances where comets can experience activity. We now point out two highly significant applications of comet activity at large distances from the Sun.

Debris disks, detected around main-sequence stars, contain a lot of small dust and ice particles. The influences of Poynting-Robertson and solar wind drugs cause a constant loss of dust particles. On the other hand, the outgassing of cometary nuclei and collision between planetesimals generate enough dust particles to sustain detected debris disks \citep{Chen2008}. The Sun has a relatively dust-poor disk compared to extrasolar debris disks. The sources of interplanetary dust particles in the outer solar system region are long-period comets and Kuiper belt objects \citep{Poppe2019}. Ice sputtering and Poynting-Robertson drugs become inefficient with the increasing distance from the Sun. As a result, distant cometary outbursts in the regions beyond the Kuiper belt can create a large amount of stable ice and dust particles. It is worth pointing out that the model of the debris disk around the Sun assumes a sharp decline in the number of Kuiper belt objects at distances > 50 AU \citep{Petit2011}. This boundary of the Kuiper Belt is probably due to observation limitations \citep{Fraser2023}. In this case, dust particles ejected from small bodies at heliocentric distances beyond 50 AU could be the indicator of comet-like objects. Based on our results, cometary outbursts in distant regions of the Solar system, where mutual collisions between small bodies are insignificant, can produce a large number of dust and ice particles. In the case of the Oort cloud, generated dust particles could periodically be forwarded to the inner parts of the Solar system during the evolution of the Oort cloud. Looking for new sources of dust particles is extremely important in light of the detection of anomalous diffuse light of unknown origin in the cosmic extragalactic background optic light \citep{Lauer2022, Carleton2022}. 

An important stage of cometary outbursts is a jet of gas and dust erupting through newly formed cracks in the comet crust. When leaving the comet surface, this jet creates the non-gravitational acceleration \citep{Marsden1973}.  Since cometary nuclei in the Oort cloud have relatively small orbital velocities, the non-gravitational acceleration can perturb the orbits of comets in addition to well-known gravitational mechanisms. In the article \citep{Belousov2023}, we examined the influence of cometary outbursts on the stability of comet orbits in the Oort cloud. We showed that cometary nuclei with a radius of less than 1 km can be expelled from the Oort cloud due to a single outburst. This effect could lead to the observed decrease in the number of the small-radius long-period comets \citep{Boe2019}.

\section{Conclusions}
\label{sec:conclusion}

We have developed the model of comet activity at large heliocentric distances caused by the recombination of radicals accumulated under cosmic ray irradiation at low temperatures. The main conclusions are the following: 
\begin{itemize}
\item Our model demonstrates that the recombination of radicals can drive cometary outbursts at large heliocentric distances and predicts comet activity even in the Oort cloud with a total ejected mass similar to the observed outbursts.
\item Recombination reactions can lead to significant chemical changes in the comet matter. 
\item The ejection of large amounts of dust and ice particles beyond the Kuiper Belt may contribute to the recently discovered anomalous diffuse light of unknown origin. 
\item Cometary outbursts in the Oort cloud may lead to the observed depletion in the number of small-radius long-period comets.    
\end{itemize} 

\section*{Acknowledgements}
This work has been supported by the grant of the Russian Science Foundation, RSF 24-22-00257. The authors want to thank Gennady Vasilyev (Ioffe Institute) for helpful comments on the local interstellar spectrum and the absorbed dose of cosmic rays.  

\section*{Declaration of Competing Interest}
The authors declare no conflicts of interest associated with this
manuscript.

\section*{Data availability}
The data underlying this article will be shared on reasonable request to the corresponding author.

\bibliographystyle{cas-model2-names}

\bibliography{MyLibrary} %

\begin{thebibliography}{41}
\expandafter\ifx\csname natexlab\endcsname\relax\def\natexlab#1{#1}\fi
\providecommand{\url}[1]{\texttt{#1}}
\providecommand{\href}[2]{#2}
\providecommand{\path}[1]{#1}
\providecommand{\DOIprefix}{doi:}
\providecommand{\ArXivprefix}{arXiv:}
\providecommand{\URLprefix}{URL: }
\providecommand{\Pubmedprefix}{pmid:}
\providecommand{\doi}[1]{\href{http://dx.doi.org/#1}{\path{#1}}}
\providecommand{\Pubmed}[1]{\href{pmid:#1}{\path{#1}}}
\providecommand{\bibinfo}[2]{#2}
\ifx\xfnm\relax \def\xfnm[#1]{\unskip,\space#1}\fi
\bibitem[{Bar-Nun et~al.(1985)Bar-Nun, Herman, Laufer and
  Rappaport}]{BarNun1985}
\bibinfo{author}{Bar-Nun, A.}, \bibinfo{author}{Herman, G.},
  \bibinfo{author}{Laufer, D.}, \bibinfo{author}{Rappaport, M.},
  \bibinfo{year}{1985}.
\newblock \bibinfo{title}{Trapping and release of gases by water ice and
  implications for icy bodies}.
\newblock \bibinfo{journal}{Icarus} \bibinfo{volume}{63},
  \bibinfo{pages}{317--332}.
\newblock \DOIprefix\doi{https://doi.org/10.1016/0019-1035(85)90048-X}.
\bibitem[{Basilevsky et~al.(2016)Basilevsky, Krasil’nikov, Shiryaev, Mall,
  Keller, Skorov, Mottola and Hviid}]{Basilevsky2016}
\bibinfo{author}{Basilevsky, A.T.}, \bibinfo{author}{Krasil’nikov, S.S.},
  \bibinfo{author}{Shiryaev, A.A.}, \bibinfo{author}{Mall, U.},
  \bibinfo{author}{Keller, H.U.}, \bibinfo{author}{Skorov, Y.V.},
  \bibinfo{author}{Mottola, S.}, \bibinfo{author}{Hviid, S.F.},
  \bibinfo{year}{2016}.
\newblock \bibinfo{title}{Estimating the strength of the nucleus material of
  comet 67p churyumov–gerasimenko}.
\newblock \bibinfo{journal}{Solar System Research} \bibinfo{volume}{50},
  \bibinfo{pages}{225--234}.
\newblock \DOIprefix\doi{10.1134/S0038094616040018}.
\bibitem[{Belousov and Pavlov(2023)}]{Belousov2023}
\bibinfo{author}{Belousov, D.V.}, \bibinfo{author}{Pavlov, A.K.},
  \bibinfo{year}{2023}.
\newblock \bibinfo{title}{Non-gravitational mechanism of comets’ ejection
  from the oort cloud due to cometary outbursts}.
\newblock \bibinfo{journal}{Solar System Research (in press)}
  \bibinfo{volume}{57}, \bibinfo{pages}{629--635}.
\bibitem[{Boe et~al.(2019)Boe, Jedicke, Meech, Wiegert, Weryk, Chambers,
  Denneau, Kaiser, Kudritzki, Magnier, Wainscoat and Waters}]{Boe2019}
\bibinfo{author}{Boe, B.}, \bibinfo{author}{Jedicke, R.},
  \bibinfo{author}{Meech, K.J.}, \bibinfo{author}{Wiegert, P.},
  \bibinfo{author}{Weryk, R.J.}, \bibinfo{author}{Chambers, K.},
  \bibinfo{author}{Denneau, L.}, \bibinfo{author}{Kaiser, N.},
  \bibinfo{author}{Kudritzki, R.P.}, \bibinfo{author}{Magnier, E.},
  \bibinfo{author}{Wainscoat, R.}, \bibinfo{author}{Waters, C.},
  \bibinfo{year}{2019}.
\newblock \bibinfo{title}{The orbit and size-frequency distribution of long
  period comets observed by pan-starrs1}.
\newblock \bibinfo{journal}{Icarus} \bibinfo{volume}{333},
  \bibinfo{pages}{252--272}.
\newblock \DOIprefix\doi{10.1016/j.icarus.2019.05.034}.
\bibitem[{Boschini et~al.(2020)Boschini, Torre, Gervasi, Grandi, Johannesson,
  Vacca, Masi, Moskalenko, Pensotti, Porter, Quadrani, Rancoita, Rozza and
  Tacconi}]{Boschini2020}
\bibinfo{author}{Boschini, M.J.}, \bibinfo{author}{Torre, S.D.},
  \bibinfo{author}{Gervasi, M.}, \bibinfo{author}{Grandi, D.},
  \bibinfo{author}{Johannesson, G.}, \bibinfo{author}{Vacca, G.L.},
  \bibinfo{author}{Masi, N.}, \bibinfo{author}{Moskalenko, I.V.},
  \bibinfo{author}{Pensotti, S.}, \bibinfo{author}{Porter, T.A.},
  \bibinfo{author}{Quadrani, L.}, \bibinfo{author}{Rancoita, P.G.},
  \bibinfo{author}{Rozza, D.}, \bibinfo{author}{Tacconi, M.},
  \bibinfo{year}{2020}.
\newblock \bibinfo{title}{Inference of the local interstellar spectra of
  cosmic-ray nuclei z < 28 with the galprop–helmod framework}.
\newblock \bibinfo{journal}{The Astrophysical Journal Supplement Series}
  \bibinfo{volume}{250}, \bibinfo{pages}{27}.
\newblock \DOIprefix\doi{10.3847/1538-4365/aba901}.
\bibitem[{Bouziani and Jewitt(2022)}]{Bouziani2022}
\bibinfo{author}{Bouziani, N.}, \bibinfo{author}{Jewitt, D.},
  \bibinfo{year}{2022}.
\newblock \bibinfo{title}{Cometary activity beyond the planets}.
\newblock \bibinfo{journal}{The Astrophysical Journal} \bibinfo{volume}{924},
  \bibinfo{pages}{37}.
\newblock \DOIprefix\doi{10.3847/1538-4357/ac323b}.
\bibitem[{Carleton et~al.(2022)Carleton, Windhorst, O’Brien, Cohen, Carter,
  Jansen, Tompkins, Arendt, Caddy, Grogin, Kenyon, Koekemoer, MacKenty,
  Casertano, Davies, Driver, Dwek, Kashlinsky, Miles, Pirzkal, Robotham, Ryan,
  Abate, Andras-Letanovszky, Berkheimer, Goisman, Henningsen, Kramer, Rogers
  and Swirbul}]{Carleton2022}
\bibinfo{author}{Carleton, T.}, \bibinfo{author}{Windhorst, R.A.},
  \bibinfo{author}{O’Brien, R.}, \bibinfo{author}{Cohen, S.H.},
  \bibinfo{author}{Carter, D.}, \bibinfo{author}{Jansen, R.},
  \bibinfo{author}{Tompkins, S.}, \bibinfo{author}{Arendt, R.G.},
  \bibinfo{author}{Caddy, S.}, \bibinfo{author}{Grogin, N.},
  \bibinfo{author}{Kenyon, S.J.}, \bibinfo{author}{Koekemoer, A.},
  \bibinfo{author}{MacKenty, J.}, \bibinfo{author}{Casertano, S.},
  \bibinfo{author}{Davies, L.J.M.}, \bibinfo{author}{Driver, S.P.},
  \bibinfo{author}{Dwek, E.}, \bibinfo{author}{Kashlinsky, A.},
  \bibinfo{author}{Miles, N.}, \bibinfo{author}{Pirzkal, N.},
  \bibinfo{author}{Robotham, A.}, \bibinfo{author}{Ryan, R.},
  \bibinfo{author}{Abate, H.}, \bibinfo{author}{Andras-Letanovszky, H.},
  \bibinfo{author}{Berkheimer, J.}, \bibinfo{author}{Goisman, Z.},
  \bibinfo{author}{Henningsen, D.}, \bibinfo{author}{Kramer, D.},
  \bibinfo{author}{Rogers, C.}, \bibinfo{author}{Swirbul, A.},
  \bibinfo{year}{2022}.
\newblock \bibinfo{title}{Skysurf: Constraints on zodiacal light and
  extragalactic background light through panchromatic hst all-sky
  surface-brightness measurements: Ii. first limits on diffuse light at 1.25,
  1.4, and 1.6 zm}.
\newblock \bibinfo{journal}{The Astronomical Journal} \bibinfo{volume}{164},
  \bibinfo{pages}{170}.
\newblock \DOIprefix\doi{10.3847/1538-3881/ac8d02}.
\bibitem[{Carpenter(1987)}]{Carpenter1987}
\bibinfo{author}{Carpenter, J.M.}, \bibinfo{year}{1987}.
\newblock \bibinfo{title}{Thermally activated release of stored chemical energy
  in cryogenic media}.
\newblock \bibinfo{journal}{Nature} \bibinfo{volume}{330},
  \bibinfo{pages}{358--360}.
\newblock \DOIprefix\doi{10.1038/330358a0}.
\bibitem[{Chen et~al.(2008)Chen, Fitzgerald and Smith}]{Chen2008}
\bibinfo{author}{Chen, C.H.}, \bibinfo{author}{Fitzgerald, M.P.},
  \bibinfo{author}{Smith, P.S.}, \bibinfo{year}{2008}.
\newblock \bibinfo{title}{A possible icy kuiper belt around hd 181327}.
\newblock \bibinfo{journal}{The Astrophysical Journal} \bibinfo{volume}{689},
  \bibinfo{pages}{539--544}.
\newblock \DOIprefix\doi{10.1086/592567}.
\bibitem[{Davidsson(2021)}]{Davidsson2021}
\bibinfo{author}{Davidsson, B.J.R.}, \bibinfo{year}{2021}.
\newblock \bibinfo{title}{Thermophysical evolution of planetesimals in the
  primordial disc}.
\newblock \bibinfo{journal}{Monthly Notices of the Royal Astronomical Society}
  \bibinfo{volume}{505}, \bibinfo{pages}{5654--5685}.
\newblock \DOIprefix\doi{10.1093/mnras/stab1593}.
\bibitem[{Flournov et~al.(1962)Flournov, Baum and Siegel}]{Flournov1962}
\bibinfo{author}{Flournov, J.M.}, \bibinfo{author}{Baum, L.H.},
  \bibinfo{author}{Siegel, S.}, \bibinfo{year}{1962}.
\newblock \bibinfo{title}{Disappearance of trapped hydrogen atoms in
  gamma-irradiated ice}.
\newblock \bibinfo{journal}{The Journal of Chemical Physics}
  \bibinfo{volume}{36}, \bibinfo{pages}{2229--2230}.
\newblock \DOIprefix\doi{10.1063/1.1732861}.
\bibitem[{Fraser et~al.(2023)Fraser, Porter, H.~W.~Lin, Kavelaars, Verbiscer,
  F.~Yoshida, Ito, Gerdes, Benecchi, Stern, Gwyn, Buie, Peltier, Singer and
  Brandy}]{Fraser2023}
\bibinfo{author}{Fraser, W.C.}, \bibinfo{author}{Porter, S.B.},
  \bibinfo{author}{H.~W.~Lin, K.~Napier, J.R.S.}, \bibinfo{author}{Kavelaars,
  J.}, \bibinfo{author}{Verbiscer, A.J.}, \bibinfo{author}{F.~Yoshida, T.T.},
  \bibinfo{author}{Ito, T.}, \bibinfo{author}{Gerdes, D.},
  \bibinfo{author}{Benecchi, S.D.}, \bibinfo{author}{Stern, S.A.},
  \bibinfo{author}{Gwyn, S.}, \bibinfo{author}{Buie, M.W.},
  \bibinfo{author}{Peltier, L.}, \bibinfo{author}{Singer, K.N.},
  \bibinfo{author}{Brandy, P.C.}, \bibinfo{year}{2023}.
\newblock \bibinfo{title}{Approaches to detecting kuiper belt objects for
  nasa’s new horizons extended mission: Digging into the noise}, in:
  \bibinfo{booktitle}{54th Lunar and Planetary Science Conference 2023}.
\bibitem[{Gerakines et~al.(1999)Gerakines, Whittet, Ehrenfreund, Boogert,
  Tielens, Schutte, Chiar, van Dishoeck, Prusti, Helmich and
  de~Graauw}]{Gerakines1999}
\bibinfo{author}{Gerakines, P.A.}, \bibinfo{author}{Whittet, D.C.B.},
  \bibinfo{author}{Ehrenfreund, P.}, \bibinfo{author}{Boogert, A.C.A.},
  \bibinfo{author}{Tielens, A.G.G.M.}, \bibinfo{author}{Schutte, W.A.},
  \bibinfo{author}{Chiar, J.E.}, \bibinfo{author}{van Dishoeck, E.F.},
  \bibinfo{author}{Prusti, T.}, \bibinfo{author}{Helmich, F.P.},
  \bibinfo{author}{de~Graauw, T.}, \bibinfo{year}{1999}.
\newblock \bibinfo{title}{Observations of solid carbon dioxide in molecular
  clouds with the infrared space observatory}.
\newblock \bibinfo{journal}{The Astrophysical Journal} \bibinfo{volume}{522},
  \bibinfo{pages}{357--377}.
\newblock \DOIprefix\doi{10.1086/307611}.
\bibitem[{Gronkowski(2007)}]{Gronkowski2007}
\bibinfo{author}{Gronkowski, P.}, \bibinfo{year}{2007}.
\newblock \bibinfo{title}{The search for a cometary outbursts mechanism: a
  comparison of various theories}.
\newblock \bibinfo{journal}{Astronomische Nachrichten} \bibinfo{volume}{328},
  \bibinfo{pages}{126--136}.
\newblock \DOIprefix\doi{10.1002/asna.200510657}.
\bibitem[{Gronkowski and Wesołowski(2015)}]{Gronkowski2015}
\bibinfo{author}{Gronkowski, P.}, \bibinfo{author}{Wesołowski, M.},
  \bibinfo{year}{2015}.
\newblock \bibinfo{title}{A model of cometary outbursts: a new simple approach
  to the classical question}.
\newblock \bibinfo{journal}{Monthly Notices of the Royal Astronomical Society}
  \bibinfo{volume}{451}, \bibinfo{pages}{3068--3077}.
\newblock \DOIprefix\doi{10.1093/mnras/stv1122}.
\bibitem[{Gronoff et~al.(2020)Gronoff, Maggiolo, Cessateur, Moore, Airapetian,
  Keyser, Dhooghe, Gibbons, Gunell, Mertens, Rubin and Hosseini}]{Gronoff2020}
\bibinfo{author}{Gronoff, G.}, \bibinfo{author}{Maggiolo, R.},
  \bibinfo{author}{Cessateur, G.}, \bibinfo{author}{Moore, W.B.},
  \bibinfo{author}{Airapetian, V.}, \bibinfo{author}{Keyser, J.D.},
  \bibinfo{author}{Dhooghe, F.}, \bibinfo{author}{Gibbons, A.},
  \bibinfo{author}{Gunell, H.}, \bibinfo{author}{Mertens, C.J.},
  \bibinfo{author}{Rubin, M.}, \bibinfo{author}{Hosseini, S.},
  \bibinfo{year}{2020}.
\newblock \bibinfo{title}{The effect of cosmic rays on cometary nuclei. i. dose
  deposition}.
\newblock \bibinfo{journal}{The Astrophysical Journal} \bibinfo{volume}{890},
  \bibinfo{pages}{89}.
\newblock \DOIprefix\doi{10.3847/1538-4357/ab67b9}.
\bibitem[{Ipatov(2012)}]{Ipatov2012}
\bibinfo{author}{Ipatov, S.I.}, \bibinfo{year}{2012}.
\newblock \bibinfo{title}{Location of upper borders of cavities containing dust
  and gas under pressure in comets}.
\newblock \bibinfo{journal}{Monthly Notices of the Royal Astronomical Society}
  \bibinfo{volume}{423}, \bibinfo{pages}{3474--3477}.
\newblock \DOIprefix\doi{10.1111/j.1365-2966.2012.21144.x}.
\bibitem[{Johnson and Quickenden(1997)}]{Johnson1997}
\bibinfo{author}{Johnson, R.E.}, \bibinfo{author}{Quickenden, T.I.},
  \bibinfo{year}{1997}.
\newblock \bibinfo{title}{Photolysis and radiolysis of water ice on outer solar
  system bodies}.
\newblock \bibinfo{journal}{Journal of Geophysical Research: Planets}
  \bibinfo{volume}{102}, \bibinfo{pages}{10985--10996}.
\newblock \DOIprefix\doi{10.1029/97JE00068}.
\bibitem[{Kaiser et~al.(1997)Kaiser, Eich, Gabrysch and Roessler}]{Kaiser1997}
\bibinfo{author}{Kaiser, R.I.}, \bibinfo{author}{Eich, G.},
  \bibinfo{author}{Gabrysch, A.}, \bibinfo{author}{Roessler, K.},
  \bibinfo{year}{1997}.
\newblock \bibinfo{title}{Theoretical and laboratory studies on the interaction
  of cosmic‐ray particles with interstellar ices. ii. formation of atomic and
  molecular hydrogen in frozen organic molecules}.
\newblock \bibinfo{journal}{The Astrophysical Journal} \bibinfo{volume}{484},
  \bibinfo{pages}{487--498}.
\newblock \DOIprefix\doi{10.1086/304316}.
\bibitem[{Kelley et~al.(2022)Kelley, Kokotanekova, Holt, Protopapa, Bodewits,
  Knight, Lister, Usher, Chatelain, Gomez, Greenstreet, Angel and
  Wooding}]{Kelley2022}
\bibinfo{author}{Kelley, M.S.P.}, \bibinfo{author}{Kokotanekova, R.},
  \bibinfo{author}{Holt, C.E.}, \bibinfo{author}{Protopapa, S.},
  \bibinfo{author}{Bodewits, D.}, \bibinfo{author}{Knight, M.M.},
  \bibinfo{author}{Lister, T.}, \bibinfo{author}{Usher, H.},
  \bibinfo{author}{Chatelain, J.}, \bibinfo{author}{Gomez, E.},
  \bibinfo{author}{Greenstreet, S.}, \bibinfo{author}{Angel, T.},
  \bibinfo{author}{Wooding, B.}, \bibinfo{year}{2022}.
\newblock \bibinfo{title}{A look at outbursts of comet c/2014 un271
  (bernardinelli–bernstein) near 20 au}.
\newblock \bibinfo{journal}{The Astrophysical Journal Letters}
  \bibinfo{volume}{933}, \bibinfo{pages}{L44}.
\newblock \DOIprefix\doi{10.3847/2041-8213/ac7bec}.
\bibitem[{Kirichek et~al.(2017)Kirichek, Lawson, Jenkins, Ridley and
  Haynes}]{Kirichek2017}
\bibinfo{author}{Kirichek, O.}, \bibinfo{author}{Lawson, C.},
  \bibinfo{author}{Jenkins, D.}, \bibinfo{author}{Ridley, C.},
  \bibinfo{author}{Haynes, D.}, \bibinfo{year}{2017}.
\newblock \bibinfo{title}{Solid methane in neutron radiation: Cryogenic
  moderators and cometary cryo volcanism}.
\newblock \bibinfo{journal}{Cryogenics} \bibinfo{volume}{88},
  \bibinfo{pages}{101--105}.
\newblock \DOIprefix\doi{10.1016/j.cryogenics.2017.10.017}.
\bibitem[{Lauer et~al.(2022)Lauer, Postman, Spencer, Weaver, Stern, Gladstone,
  Binzel, Britt, Buie, Buratti, Cheng, Grundy, Horányi, Kavelaars, Linscott,
  Lisse, McKinnon, McNutt, Moore, Núñez, Olkin, Parker, Porter, Reuter,
  Robbins, Schenk, Showalter, Singer, Verbiscer and Young}]{Lauer2022}
\bibinfo{author}{Lauer, T.R.}, \bibinfo{author}{Postman, M.},
  \bibinfo{author}{Spencer, J.R.}, \bibinfo{author}{Weaver, H.A.},
  \bibinfo{author}{Stern, S.A.}, \bibinfo{author}{Gladstone, G.R.},
  \bibinfo{author}{Binzel, R.P.}, \bibinfo{author}{Britt, D.T.},
  \bibinfo{author}{Buie, M.W.}, \bibinfo{author}{Buratti, B.J.},
  \bibinfo{author}{Cheng, A.F.}, \bibinfo{author}{Grundy, W.M.},
  \bibinfo{author}{Horányi, M.}, \bibinfo{author}{Kavelaars, J.J.},
  \bibinfo{author}{Linscott, I.R.}, \bibinfo{author}{Lisse, C.M.},
  \bibinfo{author}{McKinnon, W.B.}, \bibinfo{author}{McNutt, R.L.},
  \bibinfo{author}{Moore, J.M.}, \bibinfo{author}{Núñez, J.I.},
  \bibinfo{author}{Olkin, C.B.}, \bibinfo{author}{Parker, J.W.},
  \bibinfo{author}{Porter, S.B.}, \bibinfo{author}{Reuter, D.C.},
  \bibinfo{author}{Robbins, S.J.}, \bibinfo{author}{Schenk, P.M.},
  \bibinfo{author}{Showalter, M.R.}, \bibinfo{author}{Singer, K.N.},
  \bibinfo{author}{Verbiscer, A.J.}, \bibinfo{author}{Young, L.A.},
  \bibinfo{year}{2022}.
\newblock \bibinfo{title}{Anomalous flux in the cosmic optical background
  detected with new horizons observations}.
\newblock \bibinfo{journal}{The Astrophysical Journal Letters}
  \bibinfo{volume}{927}, \bibinfo{pages}{L8}.
\newblock \DOIprefix\doi{10.3847/2041-8213/ac573d}.
\bibitem[{Marboeuf et~al.(2012)Marboeuf, Schmitt, Petit, Mousis and
  Fray}]{Marboeuf2012}
\bibinfo{author}{Marboeuf, U.}, \bibinfo{author}{Schmitt, B.},
  \bibinfo{author}{Petit, J.M.}, \bibinfo{author}{Mousis, O.},
  \bibinfo{author}{Fray, N.}, \bibinfo{year}{2012}.
\newblock \bibinfo{title}{A cometary nucleus model taking into account all
  phase changes of water ice: amorphous, crystalline, and clathrate}.
\newblock \bibinfo{journal}{Astronomy and Astrophysics} \bibinfo{volume}{542},
  \bibinfo{pages}{A82}.
\newblock \DOIprefix\doi{10.1051/0004-6361/201118176}.
\bibitem[{Marsden et~al.(1973)Marsden, Sekanina and Yeomans}]{Marsden1973}
\bibinfo{author}{Marsden, B.G.}, \bibinfo{author}{Sekanina, Z.},
  \bibinfo{author}{Yeomans, D.K.}, \bibinfo{year}{1973}.
\newblock \bibinfo{title}{Comets and nongravitational forces. v}.
\newblock \bibinfo{journal}{The Astronomical Journal} \bibinfo{volume}{78},
  \bibinfo{pages}{211}.
\newblock \DOIprefix\doi{10.1086/111402}.
\bibitem[{Meech et~al.(2009)Meech, Pittichová, Bar-Nun, Notesco, Laufer,
  Hainaut, Lowry, Yeomans and Pitts}]{Meech2009}
\bibinfo{author}{Meech, K.}, \bibinfo{author}{Pittichová, J.},
  \bibinfo{author}{Bar-Nun, A.}, \bibinfo{author}{Notesco, G.},
  \bibinfo{author}{Laufer, D.}, \bibinfo{author}{Hainaut, O.},
  \bibinfo{author}{Lowry, S.}, \bibinfo{author}{Yeomans, D.},
  \bibinfo{author}{Pitts, M.}, \bibinfo{year}{2009}.
\newblock \bibinfo{title}{Activity of comets at large heliocentric distances
  pre-perihelion}.
\newblock \bibinfo{journal}{Icarus} \bibinfo{volume}{201},
  \bibinfo{pages}{719--739}.
\newblock \DOIprefix\doi{10.1016/j.icarus.2008.12.045}.
\bibitem[{Moore et~al.(1983)Moore, Donn, Khanna and A'Hearn}]{Moore1983}
\bibinfo{author}{Moore, M.H.}, \bibinfo{author}{Donn, B.},
  \bibinfo{author}{Khanna, R.}, \bibinfo{author}{A'Hearn, M.F.},
  \bibinfo{year}{1983}.
\newblock \bibinfo{title}{Studies of proton-irradiated cometary-type ice
  mixtures}.
\newblock \bibinfo{journal}{Icarus} \bibinfo{volume}{54},
  \bibinfo{pages}{388--405}.
\newblock \DOIprefix\doi{10.1016/0019-1035(83)90236-1}.
\bibitem[{Moore and Hudson(1992)}]{Moore1992}
\bibinfo{author}{Moore, M.H.}, \bibinfo{author}{Hudson, R.L.},
  \bibinfo{year}{1992}.
\newblock \bibinfo{title}{Far-infrared spectral studies of phase changes in
  water ice induced by proton irradiation}.
\newblock \bibinfo{journal}{The Astrophysical Journal} \bibinfo{volume}{401},
  \bibinfo{pages}{353}.
\newblock \DOIprefix\doi{10.1086/172065}.
\bibitem[{Orosei et~al.(1999)Orosei, Capaccioni, Capria, Coradini, Sanctis,
  Federico, Salomone and Huot}]{Orosei1999}
\bibinfo{author}{Orosei, R.}, \bibinfo{author}{Capaccioni, F.},
  \bibinfo{author}{Capria, M.}, \bibinfo{author}{Coradini, A.},
  \bibinfo{author}{Sanctis, M.}, \bibinfo{author}{Federico, C.},
  \bibinfo{author}{Salomone, M.}, \bibinfo{author}{Huot, J.P.},
  \bibinfo{year}{1999}.
\newblock \bibinfo{title}{Numerically improved thermochemical evolution models
  of comet nuclei}.
\newblock \bibinfo{journal}{Planetary and Space Science} \bibinfo{volume}{47},
  \bibinfo{pages}{839--853}.
\newblock \DOIprefix\doi{10.1016/S0032-0633(99)00018-5}.
\bibitem[{Pavlov et~al.(2022)Pavlov, Belousov, Tsurkov and
  Lomasov}]{Pavlov2022}
\bibinfo{author}{Pavlov, A.K.}, \bibinfo{author}{Belousov, D.V.},
  \bibinfo{author}{Tsurkov, D.A.}, \bibinfo{author}{Lomasov, V.N.},
  \bibinfo{year}{2022}.
\newblock \bibinfo{title}{Cosmic ray irradiation of comet nuclei: a possible
  source of cometary outbursts at large heliocentric distances}.
\newblock \bibinfo{journal}{Monthly Notices of the Royal Astronomical Society}
  \bibinfo{volume}{511}, \bibinfo{pages}{5909--5914}.
\newblock \DOIprefix\doi{10.1093/mnras/stac497}.
\bibitem[{Petit et~al.(2011)Petit, Kavelaars, Gladman, Jones, Parker,
  Laerhoven, Nicholson, Mars, Rousselot, Mousis, Marsden, Bieryla, Taylor,
  Ashby, Benavidez, Bagatin and Bernabeu}]{Petit2011}
\bibinfo{author}{Petit, J.M.}, \bibinfo{author}{Kavelaars, J.J.},
  \bibinfo{author}{Gladman, B.J.}, \bibinfo{author}{Jones, R.L.},
  \bibinfo{author}{Parker, J.W.}, \bibinfo{author}{Laerhoven, C.V.},
  \bibinfo{author}{Nicholson, P.}, \bibinfo{author}{Mars, G.},
  \bibinfo{author}{Rousselot, P.}, \bibinfo{author}{Mousis, O.},
  \bibinfo{author}{Marsden, B.}, \bibinfo{author}{Bieryla, A.},
  \bibinfo{author}{Taylor, M.}, \bibinfo{author}{Ashby, M.L.N.},
  \bibinfo{author}{Benavidez, P.}, \bibinfo{author}{Bagatin, A.C.},
  \bibinfo{author}{Bernabeu, G.}, \bibinfo{year}{2011}.
\newblock \bibinfo{title}{The canada-france ecliptic plane survey—full data
  release: The orbital structure of the kuiper belt}.
\newblock \bibinfo{journal}{The Astronomical Journal} \bibinfo{volume}{142},
  \bibinfo{pages}{131}.
\newblock \DOIprefix\doi{10.1088/0004-6256/142/4/131}.
\bibitem[{Poppe et~al.(2019)Poppe, Lisse, Piquette, Zemcov, Horányi, James,
  Szalay, Bernardoni and Stern}]{Poppe2019}
\bibinfo{author}{Poppe, A.R.}, \bibinfo{author}{Lisse, C.M.},
  \bibinfo{author}{Piquette, M.}, \bibinfo{author}{Zemcov, M.},
  \bibinfo{author}{Horányi, M.}, \bibinfo{author}{James, D.},
  \bibinfo{author}{Szalay, J.R.}, \bibinfo{author}{Bernardoni, E.},
  \bibinfo{author}{Stern, S.A.}, \bibinfo{year}{2019}.
\newblock \bibinfo{title}{Constraining the solar system's debris disk with in
  situ new horizons measurements from the edgeworth–kuiper belt}.
\newblock \bibinfo{journal}{The Astrophysical Journal} \bibinfo{volume}{881},
  \bibinfo{pages}{L12}.
\newblock \DOIprefix\doi{10.3847/2041-8213/ab322a}.
\bibitem[{Prialnik and Bar-Nun(1987)}]{Prialnik1987}
\bibinfo{author}{Prialnik, D.}, \bibinfo{author}{Bar-Nun, A.},
  \bibinfo{year}{1987}.
\newblock \bibinfo{title}{On the evolution and activity of cometary nuclei}.
\newblock \bibinfo{journal}{The Astrophysical Journal} \bibinfo{volume}{313},
  \bibinfo{pages}{893}.
\newblock \DOIprefix\doi{10.1086/165029}.
\bibitem[{Prialnik and Sierks(2017)}]{Prialnik2017}
\bibinfo{author}{Prialnik, D.}, \bibinfo{author}{Sierks, H.},
  \bibinfo{year}{2017}.
\newblock \bibinfo{title}{A mechanism for comet surface collapse as observed by
  rosetta on 67p/churyumov-gerasimenko}.
\newblock \bibinfo{journal}{Monthly Notices of the Royal Astronomical Society}
  \bibinfo{volume}{469}, \bibinfo{pages}{S217--S221}.
\newblock \DOIprefix\doi{10.1093/mnras/stx1577}.
\bibitem[{Reach et~al.(2010)Reach, Vaubaillon, Lisse, Holloway and
  Rho}]{Reach2010}
\bibinfo{author}{Reach, W.T.}, \bibinfo{author}{Vaubaillon, J.},
  \bibinfo{author}{Lisse, C.M.}, \bibinfo{author}{Holloway, M.},
  \bibinfo{author}{Rho, J.}, \bibinfo{year}{2010}.
\newblock \bibinfo{title}{Explosion of comet 17p/holmes as revealed by the
  spitzer space telescope}.
\newblock \bibinfo{journal}{Icarus} \bibinfo{volume}{208},
  \bibinfo{pages}{276--292}.
\newblock \DOIprefix\doi{10.1016/j.icarus.2010.01.020}.
\bibitem[{{Schmitt} et~al.(1989){Schmitt}, {Espinasse}, {Grim}, {Greenberg} and
  {Klinger}}]{Schmitt1989}
\bibinfo{author}{{Schmitt}, B.}, \bibinfo{author}{{Espinasse}, S.},
  \bibinfo{author}{{Grim}, R.J.A.}, \bibinfo{author}{{Greenberg}, J.M.},
  \bibinfo{author}{{Klinger}, J.}, \bibinfo{year}{1989}.
\newblock \bibinfo{title}{{Laboratory studies of cometary ice analogues.}}, in:
  \bibinfo{editor}{{Hunt}, J.J.}, \bibinfo{editor}{{Guyenne}, T.D.} (Eds.),
  \bibinfo{booktitle}{Physics and Mechanics of Cometary Materials}, pp.
  \bibinfo{pages}{65--69}.
\bibitem[{Shabalin(1998)}]{Shabalin1998}
\bibinfo{author}{Shabalin, E.}, \bibinfo{year}{1998}.
\newblock \bibinfo{title}{On radiation effects in water ice at low
  temperature}, in: \bibinfo{booktitle}{Proceedings of the Fourteenth Meeting
  of the International Collaboration on Advanced Neutron Sources (ICANS-XIV),
  Illinois}, \bibinfo{publisher}{Journal of Neutron Research}. pp.
  \bibinfo{pages}{497--506}.
\bibitem[{Shabalin et~al.(2003)Shabalin, Kulagin, Kulikov and
  Melikhov}]{Shabalin2003}
\bibinfo{author}{Shabalin, E.}, \bibinfo{author}{Kulagin, E.},
  \bibinfo{author}{Kulikov, S.}, \bibinfo{author}{Melikhov, V.},
  \bibinfo{year}{2003}.
\newblock \bibinfo{title}{Experimental study of spontaneous release of
  accumulated energy in irradiated ices}.
\newblock \bibinfo{journal}{Radiation Physics and Chemistry}
  \bibinfo{volume}{67}, \bibinfo{pages}{315--319}.
\newblock \DOIprefix\doi{10.1016/S0969-806X(03)00059-8}.
\bibitem[{Stern and Shull(1988)}]{Stern1988}
\bibinfo{author}{Stern, S.A.}, \bibinfo{author}{Shull, J.M.},
  \bibinfo{year}{1988}.
\newblock \bibinfo{title}{The influence of supernovae and passing stars on
  comets in the oort cloud}.
\newblock \bibinfo{journal}{Nature} \bibinfo{volume}{332},
  \bibinfo{pages}{407--411}.
\newblock \DOIprefix\doi{10.1038/332407a0}.
\bibitem[{Strazzulla and Johnson(1991)}]{Strazzulla1991}
\bibinfo{author}{Strazzulla, G.}, \bibinfo{author}{Johnson, R.E.},
  \bibinfo{year}{1991}.
\newblock \bibinfo{title}{Irradiation Effects on Comets and Cometary Debris}.
\newblock Proceedings of IAU Colloq. 116, pp. \bibinfo{pages}{243--275}.
\newblock \DOIprefix\doi{10.1007/978-94-011-3378-4_11}.
\bibitem[{Vincent et~al.(2015)Vincent, Bodewits, Besse, Sierks, Barbieri, Lamy,
  Rodrigo, Koschny, Rickman, Keller, Agarwal, A'Hearn, Auger, Barucci, Bertaux,
  Bertini, Capanna, Cremonese, Deppo, Davidsson, Debei, Cecco, El-Maarry,
  Ferri, Fornasier, Fulle, Gaskell, Giacomini, Groussin, Guilbert-Lepoutre,
  Gutierrez-Marques, Gutiérrez, Güttler, Hoekzema, Höfner, Hviid, Ip, Jorda,
  Knollenberg, Kovacs, Kramm, Kührt, Küppers, Forgia, Lara, Lazzarin, Lee,
  Leyrat, Lin, Moreno, Lowry, Magrin, Maquet, Marchi, Marzari, Massironi,
  Michalik, Moissl, Mottola, Naletto, Oklay, Pajola, Preusker, Scholten,
  Thomas, Toth and Tubiana}]{Vincent2015}
\bibinfo{author}{Vincent, J.B.}, \bibinfo{author}{Bodewits, D.},
  \bibinfo{author}{Besse, S.}, \bibinfo{author}{Sierks, H.},
  \bibinfo{author}{Barbieri, C.}, \bibinfo{author}{Lamy, P.},
  \bibinfo{author}{Rodrigo, R.}, \bibinfo{author}{Koschny, D.},
  \bibinfo{author}{Rickman, H.}, \bibinfo{author}{Keller, H.U.},
  \bibinfo{author}{Agarwal, J.}, \bibinfo{author}{A'Hearn, M.F.},
  \bibinfo{author}{Auger, A.T.}, \bibinfo{author}{Barucci, M.A.},
  \bibinfo{author}{Bertaux, J.L.}, \bibinfo{author}{Bertini, I.},
  \bibinfo{author}{Capanna, C.}, \bibinfo{author}{Cremonese, G.},
  \bibinfo{author}{Deppo, V.D.}, \bibinfo{author}{Davidsson, B.},
  \bibinfo{author}{Debei, S.}, \bibinfo{author}{Cecco, M.D.},
  \bibinfo{author}{El-Maarry, M.R.}, \bibinfo{author}{Ferri, F.},
  \bibinfo{author}{Fornasier, S.}, \bibinfo{author}{Fulle, M.},
  \bibinfo{author}{Gaskell, R.}, \bibinfo{author}{Giacomini, L.},
  \bibinfo{author}{Groussin, O.}, \bibinfo{author}{Guilbert-Lepoutre, A.},
  \bibinfo{author}{Gutierrez-Marques, P.}, \bibinfo{author}{Gutiérrez, P.J.},
  \bibinfo{author}{Güttler, C.}, \bibinfo{author}{Hoekzema, N.},
  \bibinfo{author}{Höfner, S.}, \bibinfo{author}{Hviid, S.F.},
  \bibinfo{author}{Ip, W.H.}, \bibinfo{author}{Jorda, L.},
  \bibinfo{author}{Knollenberg, J.}, \bibinfo{author}{Kovacs, G.},
  \bibinfo{author}{Kramm, R.}, \bibinfo{author}{Kührt, E.},
  \bibinfo{author}{Küppers, M.}, \bibinfo{author}{Forgia, F.L.},
  \bibinfo{author}{Lara, L.M.}, \bibinfo{author}{Lazzarin, M.},
  \bibinfo{author}{Lee, V.}, \bibinfo{author}{Leyrat, C.},
  \bibinfo{author}{Lin, Z.Y.}, \bibinfo{author}{Moreno, J.J.L.},
  \bibinfo{author}{Lowry, S.}, \bibinfo{author}{Magrin, S.},
  \bibinfo{author}{Maquet, L.}, \bibinfo{author}{Marchi, S.},
  \bibinfo{author}{Marzari, F.}, \bibinfo{author}{Massironi, M.},
  \bibinfo{author}{Michalik, H.}, \bibinfo{author}{Moissl, R.},
  \bibinfo{author}{Mottola, S.}, \bibinfo{author}{Naletto, G.},
  \bibinfo{author}{Oklay, N.}, \bibinfo{author}{Pajola, M.},
  \bibinfo{author}{Preusker, F.}, \bibinfo{author}{Scholten, F.},
  \bibinfo{author}{Thomas, N.}, \bibinfo{author}{Toth, I.},
  \bibinfo{author}{Tubiana, C.}, \bibinfo{year}{2015}.
\newblock \bibinfo{title}{Large heterogeneities in comet 67p as revealed by
  active pits from sinkhole collapse}.
\newblock \bibinfo{journal}{Nature} \bibinfo{volume}{523},
  \bibinfo{pages}{63--66}.
\newblock \DOIprefix\doi{10.1038/nature14564}.
\bibitem[{Zhu et~al.(2021)Zhu, Bergantini, Singh, Abplanalp and
  Kaiser}]{Zhu2021}
\bibinfo{author}{Zhu, C.}, \bibinfo{author}{Bergantini, A.},
  \bibinfo{author}{Singh, S.K.}, \bibinfo{author}{Abplanalp, M.J.},
  \bibinfo{author}{Kaiser, R.I.}, \bibinfo{year}{2021}.
\newblock \bibinfo{title}{Rapid radical–radical induced explosive desorption
  of ice-coated interstellar nanoparticles}.
\newblock \bibinfo{journal}{The Astrophysical Journal} \bibinfo{volume}{920},
  \bibinfo{pages}{73}.
\newblock \DOIprefix\doi{10.3847/1538-4357/ac116a}.

\end{thebibliography}

\vskip3pt

\end{document}